\documentclass[11pt,english]{article}
\pdfoutput=1
\usepackage{lmodern}
\usepackage[ansinew]{inputenc}
\usepackage{geometry}
\usepackage{amsmath}
\usepackage{amssymb}
\usepackage{esint}

\makeatletter
\usepackage{hyperref}
\usepackage{cite}
\usepackage{babel}
\usepackage{mathdots}
\usepackage{amsfonts}
\usepackage{mathrsfs}
\usepackage{amsmath}
\usepackage{amssymb}
\usepackage{esint}
\usepackage{graphicx}
\usepackage{youngtab}
\usepackage{multicol}
\usepackage{multirow}
\usepackage{slashed}
\usepackage{bbm}
\usepackage[dvipsnames]{xcolor}
\usepackage[normalem]{ulem}

\usepackage{feynmp-auto}
\allowdisplaybreaks

\makeatother

\begin{document}
\vspace*{-2.5cm}
\begin{flushright}
{\small
LMU-ASC 72/15\\
MPP-2015-275
}
\end{flushright}
\renewcommand{\thefootnote}{\fnsymbol{footnote}}
\vspace{0.5cm}
\begin{center}
{\LARGE \bf   Stringy explanation of $b\to s \ell^+ \ell^-$ anomalies} \\
\bigskip\color{black}\vspace{0.6cm}
{{\large \bf Alejandro Celis$^{a}$, Wan-Zhe~Feng$^{a,b}$ and Dieter~L\"ust$^{a,b}$\footnote{Email: \url{alejandro.celis@physik.uni-muenchen.de},~ \url{vicf@mpp.mpg.de},~ \url{dieter.luest@lmu.de}}}
} \\[7mm]
{\it  (a)  Arnold Sommerfeld Center for Theoretical Physics, Fakult\"at f\"ur Physik, \\  Ludwig-Maximilians-Universit\"at M\"unchen,  80333 M\"unchen, Germany}\\[3mm]
{\it  (b)  Max--Planck--Institut f\"ur Physik (Werner--Heisenberg--Institut), 80805 M\"unchen, Germany}\\[3mm]
\end{center}
\bigskip
\bigskip
\bigskip
\centerline{\large\bf Abstract}
\begin{quote}
  We show that the recent anomalies in $b \rightarrow s \ell^+ \ell^-$ transitions observed by the LHCb collaboration
  can be accommodated within string motivated models with a low mass $Z^{\prime}$ gauge boson.
  Such $Z^{\prime}$ gauge boson can be obtained in compactifications with a low string scale.
  We consider a class of intersecting D-brane models
  in which different families of quarks and leptons are simultaneously realized at different D-brane intersections.
  The explanation of $b \rightarrow s \ell^+ \ell^-$ anomalies via a stringy $Z^{\prime}$
  sets important restrictions on these viable D-brane constructions.
\end{quote}

\renewcommand{\thefootnote}{\arabic{footnote}}\setcounter{footnote}{0}

\newpage

\section{Introduction}

Current data for $b \rightarrow s \ell^+ \ell^-$ decays shows a series of deviations from the Standard Model (SM)~\cite{Aaij:2013qta,Aaij:2015esa,Aaij:2014ora}.     Among the relevant observables, the ratio $R_K$  measuring the $B^+ \rightarrow K^+ \mu^+ \mu^-$ rate normalized by the electron mode is particularly interesting since it is known within the SM with a very good accuracy and constitutes a test of lepton universality in $B$-meson decays~\cite{Hiller:2003js}.     This ratio has been measured recently by the LHCb collaboration in the $q^2$-range $1<q^2< 6$~GeV$^2$ giving $R_K =0.745^{+0.090}_{-0.074}\pm0.036$, representing a $2.6\sigma$ deviation from the SM prediction $R_K \simeq 1$~\cite{Aaij:2014ora}.     The measurement of the observable $P_5^{\prime}$ in $B^0 \rightarrow K^{*} \mu^+ \mu^-$ decays shows a tension with respect to the SM at the $3.7\sigma$ level~\cite{Aaij:2013qta,Descotes-Genon:2013vna}.  Additionally, the differential branching fraction $B_s \rightarrow  \phi \mu^+ \mu^-$ was recently reported to be $3.5\sigma$ below the SM prediction in the low-$q^2$ region~\cite{Aaij:2015esa}.

\begin{figure}[ht!]
\begin{center}
\begin{fmffile}{Zvertex}
    \begin{fmfgraph*}(30,20)
        \fmfright{o1,o2,o3}   \fmflabel{$s$}{o1}
        \fmfleft{i}  \fmflabel{$b$}{i}
        \fmf{fermion}{i,v1,o1}
         \fmf{photon}{v1,v2}    \fmf{photon,label=$Z^{\prime}$}{v1,v2}
        \fmf{fermion}{o2,v2,o3}    \fmflabel{$\ell^-$}{o3}    \fmflabel{$\ell^+$}{o2}
    \end{fmfgraph*}
\end{fmffile}
\end{center}
\caption{\textit{Explanation of the $b \rightarrow s \ell^+ \ell^-$ anomalies due to the tree-level exchange of a heavy $Z^{\prime}$ boson.   } } \label{fig:basicf}
\end{figure}
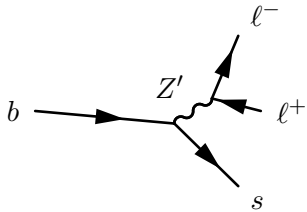

A massive $Z^{\prime}$ boson of mass of $\mathcal{O}(10)$~TeV
is a possible new physics candidate to explain the observed $b\to s \ell^+ \ell^-$ anomalies~\cite{Gauld:2013qja,Buras:2013qja,Gauld:2013qba,Buras:2013dea,Altmannshofer:2014cfa,Crivellin:2015mga,Crivellin:2015lwa,Sierra:2015fma,Celis:2015ara,Belanger:2015nma,Altmannshofer:2015mqa,Falkowski:2015zwa,Allanach:2015gkd,Niehoff:2015bfa,Niehoff:2015iaa,Carmona:2015ena}, see Figure~\ref{fig:basicf}.         A neutral boson at the TeV scale can evade current limits from the LHC while potentially produce the relevant deviations from the SM in $b\to s \ell^+ \ell^-$ transitions if the following conditions are met:
\begin{itemize}
\item Flavor changing $Z_{\alpha}^{\prime} \bar s_L  \gamma^{\alpha}    b_L $ couplings are present at tree-level.
\item  The $Z^{\prime}$ boson couples differently to muons and electrons in order to accommodate $R_K$.
\end{itemize}
%

A $Z^{\prime}$ boson with the required characteristics could arise as a massive abelian gauge boson associated to an anomaly free $\mathrm{U(1)}^{\prime}$ symmetry which is spontaneously broken around the TeV scale, obtaining its mass via the Higgs mechanism.      The non-universal character of the $Z^{\prime}$ couplings to different families
can be traced in these models to fermion mixing effects or to horizontal $\mathrm{U(1)}^{\prime}$ charges~\cite{Gauld:2013qja,Buras:2013qja,Gauld:2013qba,Buras:2013dea,Altmannshofer:2014cfa,Crivellin:2015mga,Crivellin:2015lwa,Sierra:2015fma,Celis:2015ara,Belanger:2015nma,Altmannshofer:2015mqa,Falkowski:2015zwa,Allanach:2015gkd}.   Another possibility would be a vector resonance from a strong dynamics associated to the electroweak symmetry breaking mechanism~\cite{Niehoff:2015bfa,Niehoff:2015iaa,Carmona:2015ena}.

We propose here a new, stringy solution to the $b \rightarrow s \ell^+ \ell^-$ puzzles.
We consider string inspired abelian gauge symmetries from intersecting D-brane models (for reviews see~\cite{Blumenhagen:2005mu,Blumenhagen:2006ci}).
These $\mathrm{U(1)}^{\prime}$s suffer from mixed anomalies with the SM which are cancelled by the Green-Schwarz mechanism~\cite{Green:1984sg}.
The corresponding gauge boson typically acquires a string scale mass
via the coupling between the two-index antisymmetric fields from the closed string sector.
Assuming the string scale is as low as TeVs,\footnote{
Due to the form of the $\mathrm{U(1)}$ mass-squared matrix Eq.~\eqref{MASSM}
and also the repulsion effect of the eigenvalues of a positive-definite matrix,
for a $Z'$ around a few TeVs,
the corresponding string scale can usually be several orders higher~\cite{Feng:2014eja,Feng:2014cla}.}
these $\mathrm{U(1)}^{\prime}$ gauge bosons can give rise to a rich phenomenology~\cite{Ghilencea:2002da,Langacker:2008yv,Berenstein:2008xg,Kumar:2007zza,Antoniadis:2009ze,Williams:2011qb,Anchordoqui:2011ag,Anchordoqui:2011eg,Feng:2014eja,Feng:2014cla,Anchordoqui:2015uea}.
As discussed in \cite{Anchordoqui:2015uea}, they also could be the stringy origin of possible diboson and dijet excesses at the LHC.
The required non-universal character of the $Z^{\prime}$ to solve the  $b \rightarrow s \ell^+ \ell^-$ puzzles
would be ultimately related in this case to geometric properties of the intersecting D-brane model.

This paper is organized as follows. In Section~\ref{sec:san} we will discuss the implications of the $b \rightarrow s \ell^+ \ell^-$ anomalies for a stringy $Z^{\prime}$ without entering into details of the D-brane construction considered.
We discuss generic aspects of the intersecting D-brane models in Section~\ref{eqmasmix},
especially, additional $\mathrm{U(1)}$'s in D-brane models and the mass mixing effect among them.
In Section~\ref{sec:anomaM} we describe the class of intersecting D-brane models considered for our analysis.    We conclude in Section~\ref{conc}.

\section{Stringy solution to the $b \rightarrow s \ell^+ \ell^-$ puzzles  \label{sec:san}}

The main finding of our work is that a class of intersecting D-brane models can explain current $b \rightarrow s \ell^+ \ell^-$ anomalies.
The model we will use for illustration is a toroidal model.\footnote{
In a purely toroidal intersecting D-brane model it is in general not possible to obtain a low string scale, but the D-branes
have to be embedded in a different space, which allows for a large transversal space (see also some comments later in the paper).}
In the so-called $\rm{U(2)}$-type models~\cite{Ibanez:2001nd},
left-handed quarks arise from different D-brane intersections and can give rise to the desired $Z_{\alpha}^{\prime} \bar s_L  \gamma^{\alpha}    b_L $ coupling due to quark mixing effects.
In addition, a subclass of models within this category, five-stack models studied in~\cite{Kokorelis:2002zz},\footnote{
To explain the $b \to s \ell^+ \ell^-$ anomalies within four-stack D-brane models
(requiring different families of quarks and leptons simultaneously realized at different brane intersections),
it seems necessary to add chiral exotics in the particle spectrum. We will not discuss this possibility here.}
also realize lepton generations at different D-brane intersections and generate non-universal $Z'$ couplings to leptons.
These features allow for the potential explanation of $b \rightarrow s \ell^+ \ell^-$ anomalies in terms of a stringy $Z^{\prime}$.

\subsection{$b \rightarrow s \ell^+ \ell^-$ anomalies}

New physics contributions to semi-leptonic $b \rightarrow s \ell^+ \ell^-$ decays are described in a model-independent manner using the effective weak Hamiltonian~\cite{Buras:1998raa}
\begin{equation}
 \mathcal{H}_{\mbox{\scriptsize{eff}}} \supset   - \frac{4 G_F }{  \sqrt{2}}   \frac{\alpha}{ 4 \pi}  V_{ts}^* V_{tb} \sum_{i} \left[   C^{\ell}_{i}(\mu)  Q^{\ell}_{i}(\mu) + C_{i}^{\prime \ell}(\mu)  Q_{i}^{\prime \ell}(\mu) \right]  \,.
\end{equation}
We focus here on the operators
\begin{align}
Q_9^{\ell} &= ( \bar s \gamma_{\mu}  P_L b  ) (   \bar \ell \gamma^{\mu}  \ell )  \,,  \qquad Q_{10}^{\ell} = ( \bar s \gamma_{\mu}  P_L b  ) (   \bar \ell \gamma^{\mu} \gamma_{5}  \ell )  \,, \nonumber \\
Q_9^{\prime \ell} &= ( \bar s \gamma_{\mu}  P_R b  ) (   \bar \ell \gamma^{\mu}  \ell )  \,,  \qquad Q_{10}^{\prime \ell} = ( \bar s \gamma_{\mu}  P_R b  ) (   \bar \ell \gamma^{\mu} \gamma_{5}  \ell )  \,.
\end{align}
In the SM the Wilson coefficients are $C_9^{\mbox{\scriptsize{SM}} \ell}   \simeq -  C_{10}^{\mbox{\scriptsize{SM}}\ell } \simeq 4.2$ at the $m_b$ scale,  while $C_{9,10}^{ \prime  \ell}\simeq 0$.   We write in the following $C^{\ell}_{i} = C^{\mbox{\scriptsize{SM}} \ell}_{i} + C^{\mbox{\scriptsize{NP}} \ell}_{i}   $.

Global fits of the Wilson coefficients to the available $b \rightarrow s \ell^+ \ell^-$ data have been performed by different groups, with varying statistical methods and treatments of hadronic uncertainties~\cite{Descotes-Genon:2015uva,Altmannshofer:2014rta,Altmannshofer:2015sma,Ghosh:2014awa,Beaujean:2013soa,Hurth:2014vma,Jager:2014rwa,Khodjamirian:2010vf,Khodjamirian:2012rm}.     Notably, the observed pattern of deviations in $b \rightarrow s \ell^+ \ell^-$ transitions seems to hint towards a consistent interpretation in terms of new physics.     Current data favors new physics contribution in $C_9^{\mbox{\scriptsize{NP}} \mu}$ of size $C_9^{\mbox{\scriptsize{NP}} \mu} \sim -1$~\cite{Descotes-Genon:2015uva,Altmannshofer:2015sma,Ghosh:2014awa,Beaujean:2013soa,Hurth:2014vma,Jager:2014rwa}.  The reported significance of this scenario varies within the different analyses available, a recent comprehensive study quotes a significance of $\sim4 \sigma$~\cite{Descotes-Genon:2015uva}.

\subsection{Implications of $b \rightarrow s \ell^+ \ell^-$ data}
We describe the main phenomenological results of our work in this section, leaving technical details of the intersecting D-brane models for the following sections.  The $Z^{\prime}$ couplings to fermions can be parametrized as
\begin{equation} \label{eqlagz}
\mathcal{L}_{{\rm NC}}^{Z'}=Z'_{\mu}\sum_{f}\big[g_{L}^{f}\bar{f}_{L}\gamma^{\mu}f_{L}+g_{R}^{f}\bar{f}_{R}\gamma^{\mu}f_{R}\big]\,.
\end{equation}
As will be explained in Section~\ref{sec:anomaM},  $q$ and $Q$ ($l$ and $\ell$) denote generations of left-handed quarks (leptons) arising from different D-brane intersections.
We identify $q$ to be $(u,d)_L,(c,s)_L$, and $Q$ to be $(t,b)_L$;
we also identify $l$ to be $(e,\nu_e)_L$ and $\ell$ to be $(\mu,\nu_\mu)_L,(\tau,\nu_\tau)_L$.
The three generations of right-handed quarks are realized at the same intersections, $U,D$ represent $(u,c,t)_R$ and $(d,s,b)_R$ respectively.
Right-handed leptons are realized at different brane intersections,
$(e,\nu)_R$ is identified with the right-handed electron and electron-neutrino while $(E, N)_R$ are identified with the second and third generation right-handed leptons.
The couplings $g_{L,R}^{f}$ are determined by a limited number of free variables within our underlying intersecting D-brane model and are subject to specific correlations.     In the fermion mass basis~\cite{Barger:2003hg,Barger:2009qs}
\begin{equation}
\mathcal{L}_{{\rm NC}}^{Z'} = Z'_{\mu}\sum_{f} \sum_{i,j} \big[B_{ij}^{f_{L}} \,  \bar{f}_{iL}\gamma^{\mu}f_{jL}+ B_{ij}^{f_R} \, \bar{f}_{iR}\gamma^{\mu}f_{jR}\big]\,,
\end{equation}
with
\begin{equation}
B^{f_L} = V_{f_L}^{\dag} \, \mathrm{diag}(g_{L}^{f_1},g_{L}^{f_2},g_{L}^{f_3})  \, V_{f_L} \,, \qquad B^{f_R} = V_{f_R}^{\dag} \, \mathrm{diag}(g_{R}^{f_1},g_{R}^{f_2},g_{R}^{f_3}) \,  V_{f_R} \,.
\end{equation}
The unitary matrices $V_{f_{L,R}}$ satisfy $V_{uL}^{\dag} V_{dL} = V_{\mbox{\scriptsize{CKM}}}$, with $V_{\mbox{\scriptsize{CKM}}}$ being the CKM matrix.
The form of these unitary matrices depend on the Higgs sector of the model and the allowed Yukawa couplings due to the new gauge symmetries.
In our model $g_R^{q_i}$ are family universal so that $B^{q_R}$ is flavor diagonal for up-and down-type quarks.    On the other hand, the left-handed couplings are non-universal, giving rise to left-handed flavor changing $Z^{\prime}$ couplings to quarks.  We focus here on the down-quark sector.
We can parametrize generically the left-handed down-quark rotation matrix as
\begin{equation}
V_{d_{L}}  = \begin{pmatrix}
 W_{d_{L}}  &  X_{d_{L}}  \\
Y_{d_{L}}  &    Z_{d_{L}}   \\
\end{pmatrix} \,,
\end{equation}
where $W_{d_{L}}$ is a $2\times 2$ sub-matrix.     The matrix $B^{d_{L}}$ can be written as
\begin{equation}
B^{d_{L}}  = \begin{pmatrix}
g_{L}^{q}  W_{d_L}^{\dag}  W_{d_L} + g_{L}^{Q} Y_{d_L}^{\dag} Y_{d_L}        &  g_{L}^{q}  W_{d_L}^{\dag}  X_{d_L} +   g_{L}^{Q} Y_{d_L}^{\dag} Z_{d_L}         \\[0.3cm]
g_{L}^{q}  X_{d_L}^{\dag}  W_{d_L} + g_{L}^{Q} Z_{d_L}^{\dag} Y_{d_L}   &   g_{L}^{q}  X_{d_L}^{\dag}  X_{d_L} + g_{L}^{Q} Z_{d_L}^{\dag} Z_{d_L}        \\
\end{pmatrix} \,.
\label{BdL}
\end{equation}
We assume that the elements $X_{d_{L}}, Y_{d_{L}}$ are small.    Taking into account the unitarity of $V_{d_L}$ we can approximate
\begin{equation} \label{eqBq}
B^{d_{L}}  = \begin{pmatrix}
g_{L}^{q} \times \mathbf{1}_{2\times2}   &  ( g_{L}^{q}  -  g_{L}^{Q} ) W_{d_L}^{\dag}  X_{d_L}        \\[0.3cm]
(g_{L}^{q} - g_{L}^{Q} )  X_{d_L}^{\dag}  W_{d_L}  &    g_{L}^{Q}     \\
\end{pmatrix} \,.
\end{equation}
Note that the $b$-$d$ and $b$-$s$ $Z^{\prime}$ couplings are unrelated a priori since they depend on unknown matrix elements of $V_{d_{L}}$, we will focus here on the $B_{sb}^{d_{L}}$ coupling.     For the charged leptons our model gives
\begin{equation}
B^{\ell_{L}}  = \begin{pmatrix}
g_{L}^{l}         &   0   \\[0.3cm]
 0  &   g_{L}^{\ell}  \times \mathbf{1}_{2\times2}       \\
\end{pmatrix} \,,
\qquad  \quad
B^{\ell_{R}}  = \begin{pmatrix}
g_{R}^{e}         & 0      \\[0.3cm]
0   &   g_{R}^{E}  \times \mathbf{1}_{2\times2}       \\
\end{pmatrix} \,.
\label{Bl}
\end{equation}
%
%

\begin{figure}[!h]
\begin{center}
~
 \includegraphics[width=5cm]{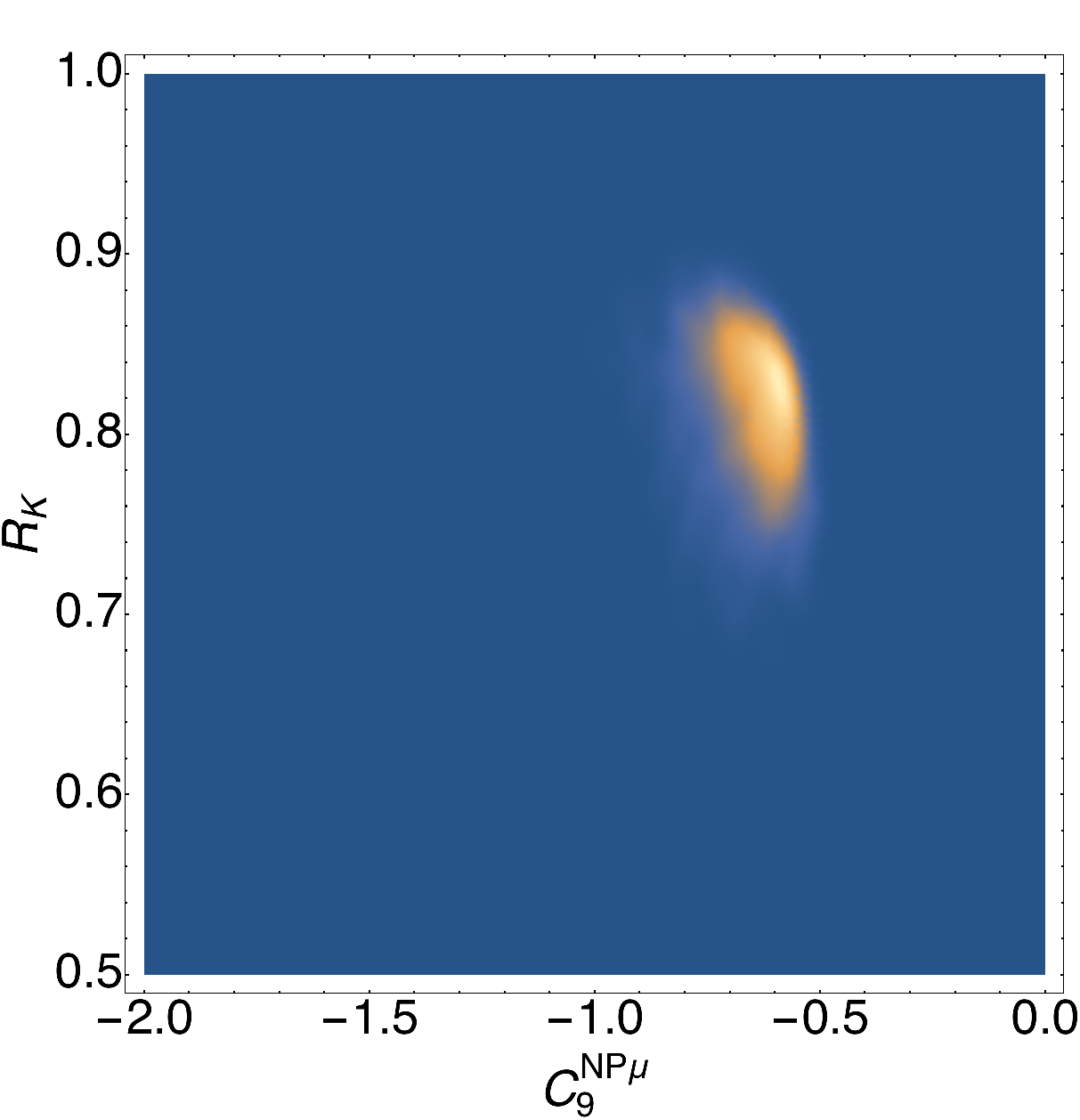}
 ~
 \includegraphics[width=5.2cm]{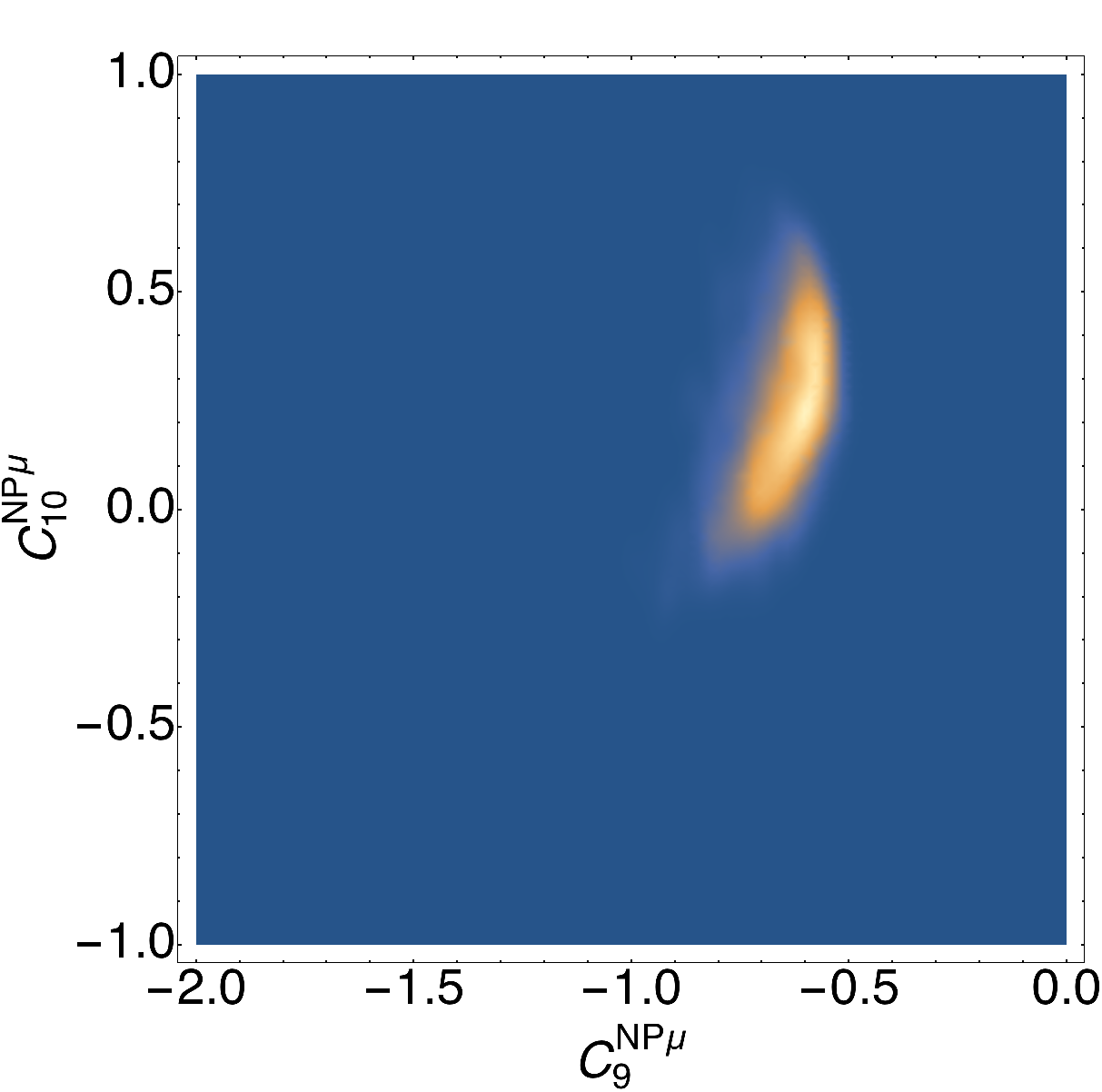}\\
 \includegraphics[width=5.2cm]{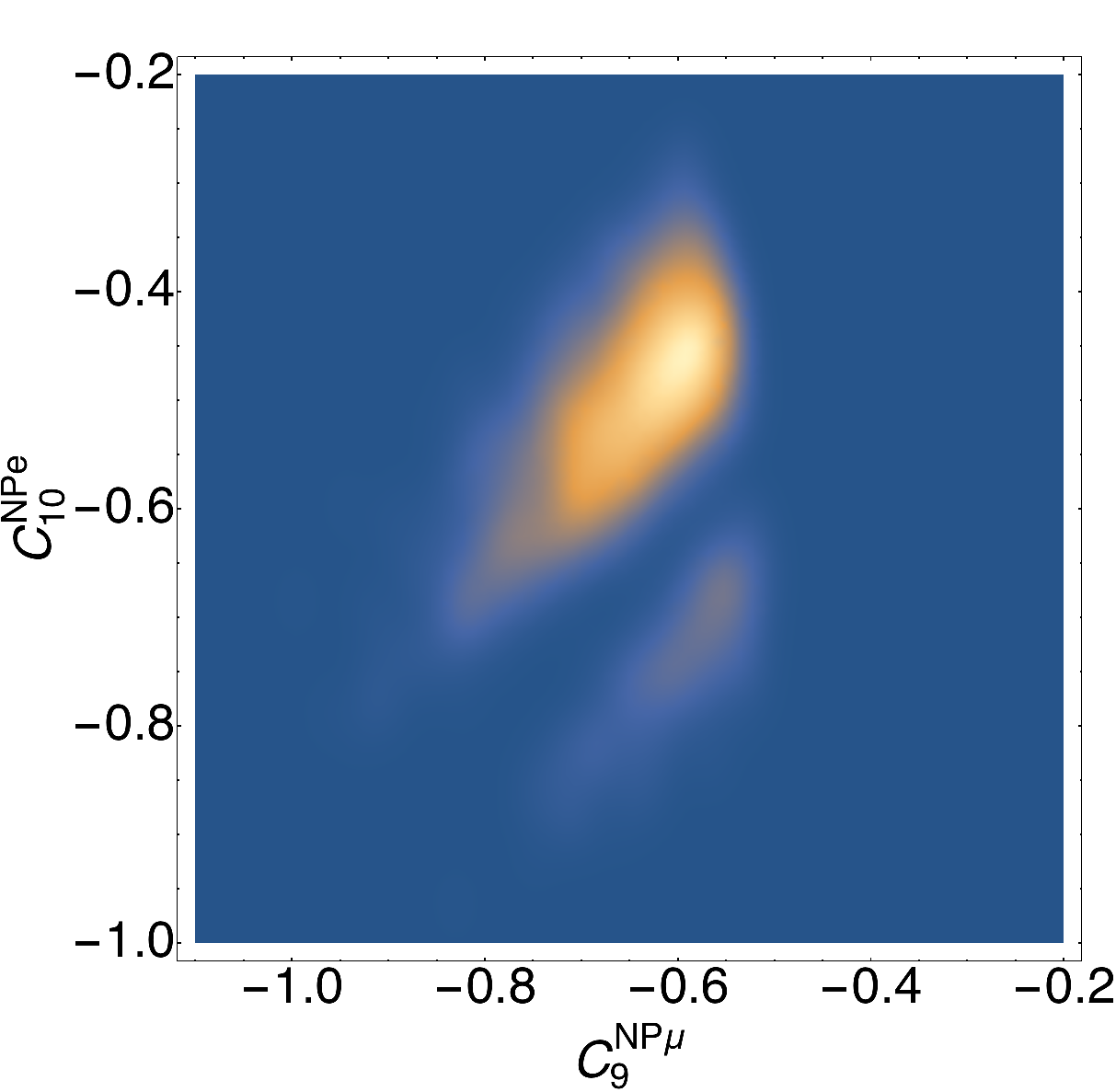}
 ~
  \includegraphics[width=5.2cm]{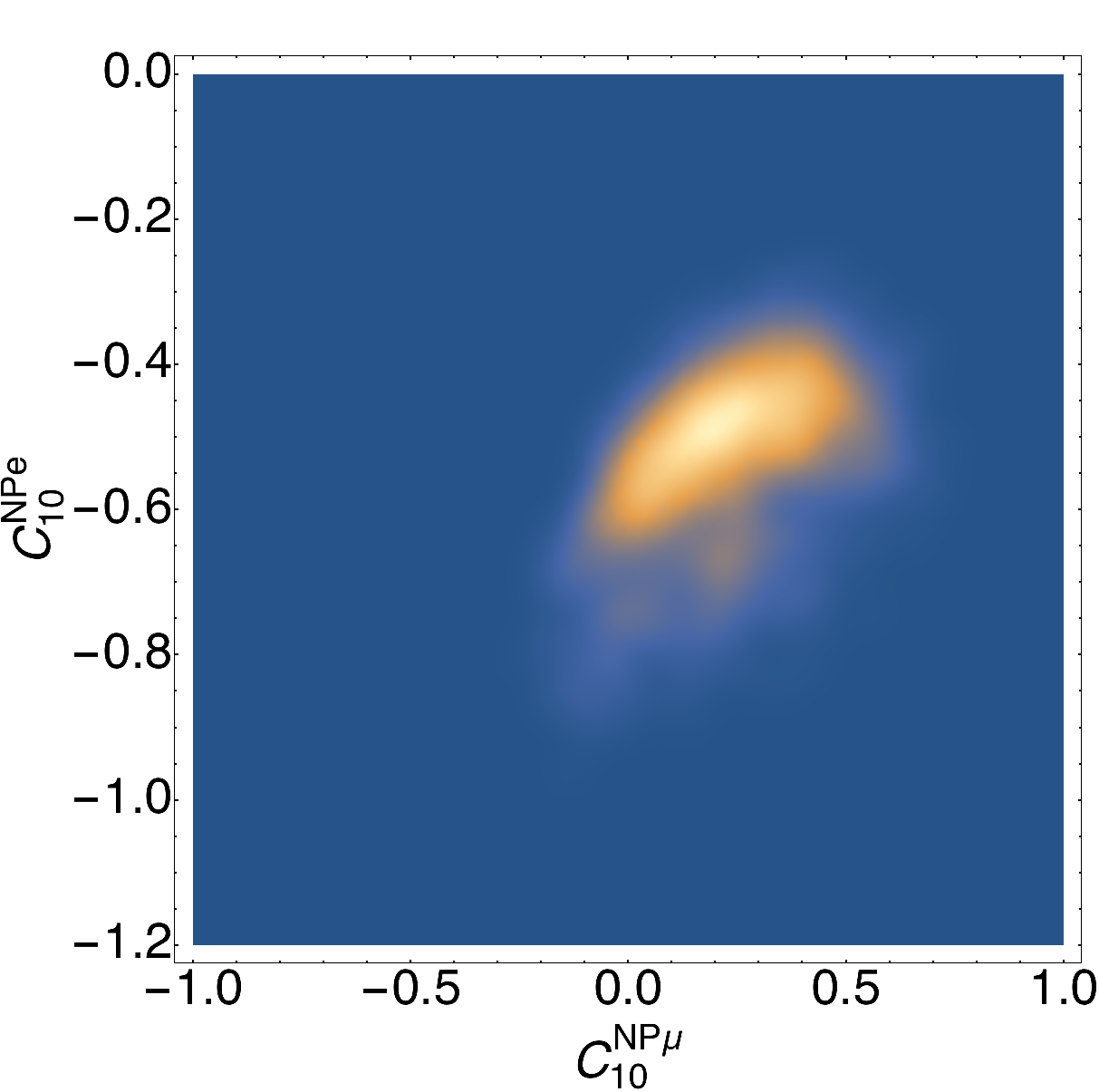}
 \end{center}
 \caption{\textit{Constraints on the Wilson coefficients from $Z^{\prime}$ searches at the LHC,  $B_s$-meson mixing, $b \rightarrow s \mu^+ \mu^-$ data and the $R_K$ measurement at $95\%$~CL.      \label{figWC}}}
\end{figure}

\begin{figure}[!h]
\begin{center}
 \includegraphics[width=5cm]{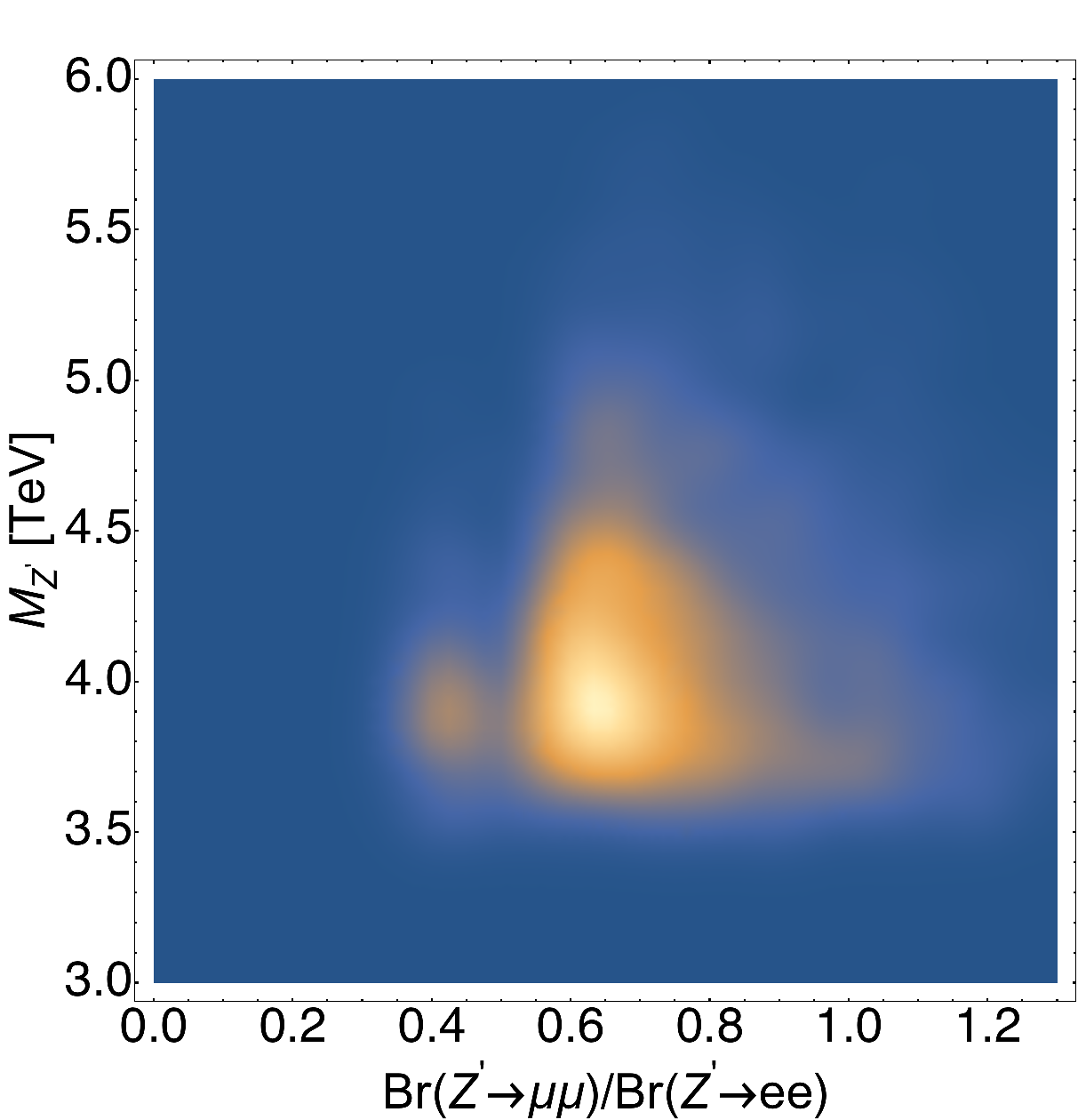}
   ~
  \includegraphics[width=5cm]{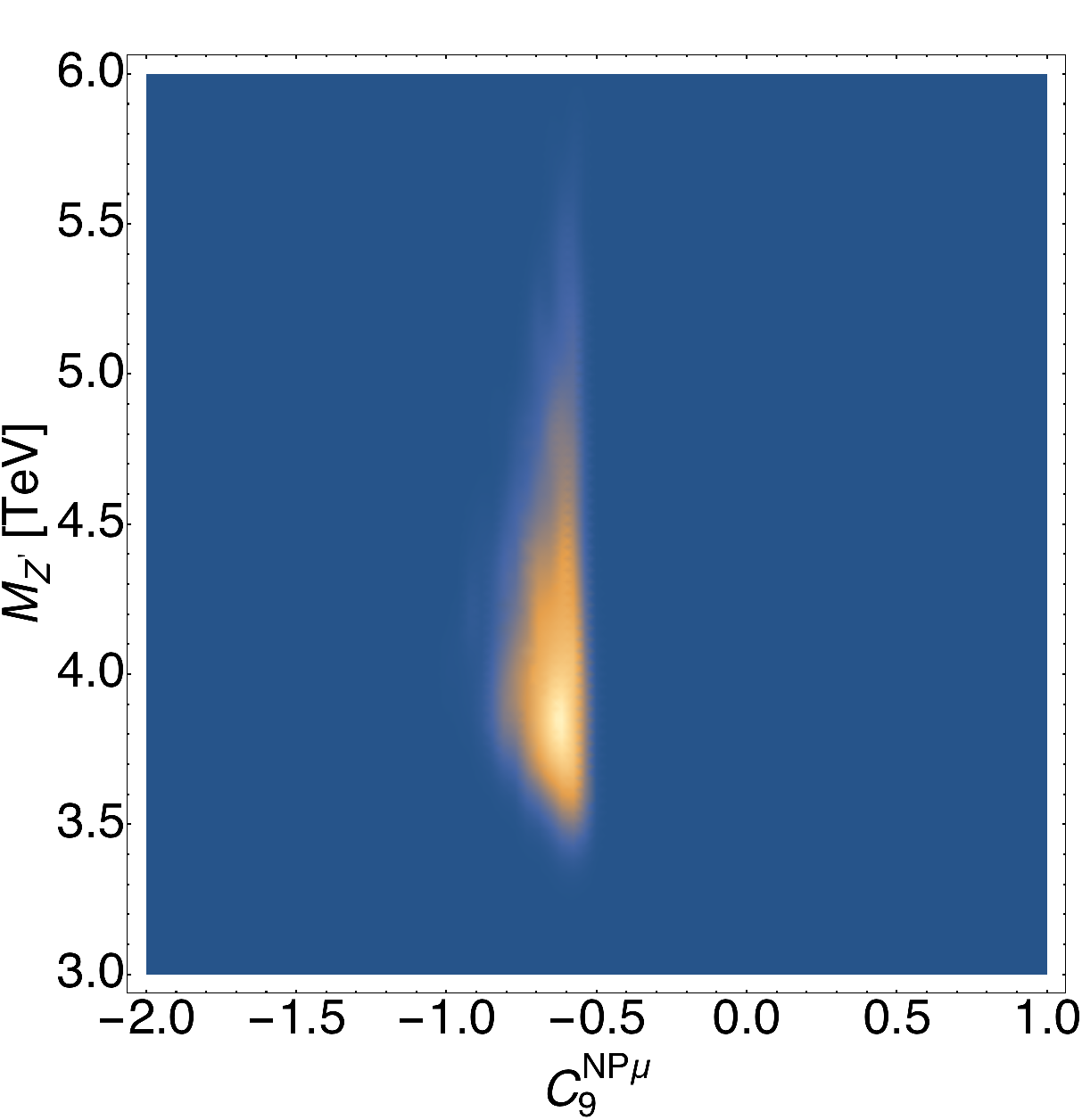}
  ~
  \includegraphics[width=5cm]{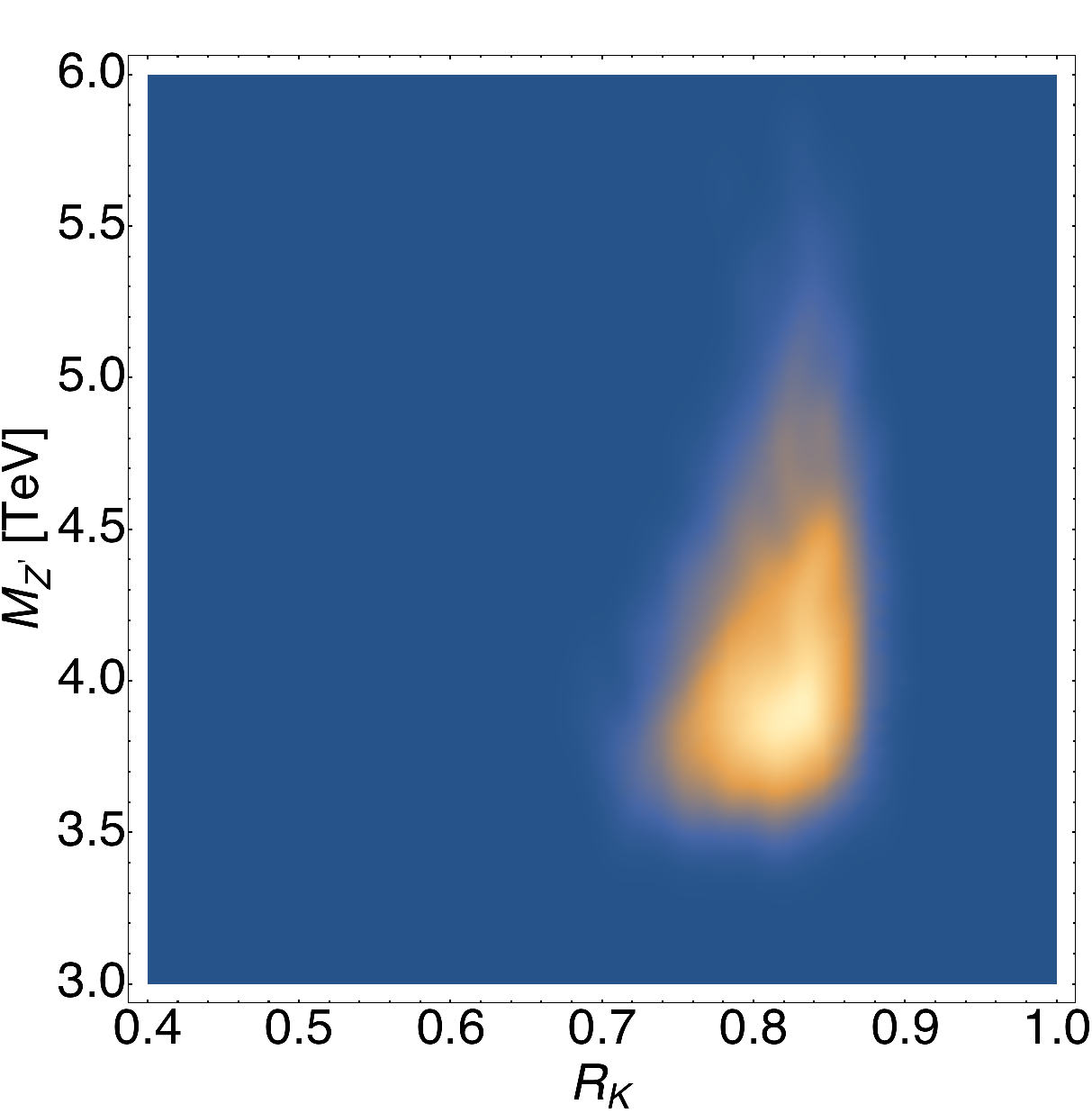}
 \end{center}
  \caption{\textit{Bounds derived from $Z^{\prime}$ searches at the LHC, $B_s$-meson mixing, $b \rightarrow s \mu^+ \mu^-$ data and the $R_K$ measurement at $95\%$~CL.      \label{figtu} }}
\end{figure}

The $Z^{\prime}$ contributions to the Wilson coefficients $C_{9,10}^{\ell}$ are given by~\cite{Buras:2012jb}
\begin{align}
C_9^{\mbox{\scriptsize{NP}} \ell} &= -  \frac{    \pi  }{   \sqrt{2}    \alpha   G_F } \frac{1}{M_{Z^{\prime}}^2}   \frac{    B_{sb}^{d_{L}}    (B_{\ell \ell}^{e_{R}} +  B_{\ell \ell}^{e_{L}} )     }{    V_{ts}^*  V_{tb} } \,, \qquad
C_{10}^{\mbox{\scriptsize{NP}} \ell} = -  \frac{    \pi  }{   \sqrt{2}    \alpha   G_F } \frac{1}{M_{Z^{\prime}}^2}   \frac{    B_{sb}^{d_{L}}  ( B_{\ell \ell}^{e_{R}} -   B_{\ell \ell}^{e_{L}} )     }{    V_{ts}^*  V_{tb} }  \,.
\end{align}
The Wilson coefficients $C_{9,10}^{\mbox{\scriptsize{NP}} \prime  \ell} $ do not receive $Z^{\prime}$ contributions at tree-level due to the absence of right-handed flavor changing neutral currents (FCNCs).  The vectorial $Z^{\prime}$ coupling to muons and electrons happen to be the same in our model, $g_{L}^{l}  + g_{R}^{e}      = g_{L}^{\ell}  +  g_{R}^{E}$, which implies that
\begin{equation}
C_{9}^{\mbox{\scriptsize{NP}} \mu} = C_{9}^{\mbox{\scriptsize{NP}} e}  \,.
\end{equation}
This correlation has important implications given that current data favors sizable new physics contributions to $C_{9}^{\mu}$.  The ratio $R_K$ defined via
\begin{equation}
R_K  =  \frac{ \displaystyle\int^{q^2_{\mbox{\scriptsize{max}}}}_{q^2_{\mbox{\scriptsize{min}}}}      \dfrac{d\Gamma(B^+ \rightarrow K^+ \mu^+ \mu^-)}{  dq^2}  dq^2  }{    \displaystyle\int^{q^2_{\mbox{\scriptsize{max}}}}_{q^2_{\mbox{\scriptsize{min}}}}      \dfrac{d\Gamma(B^+ \rightarrow K^+ e^+ e^-)}{  dq^2}  dq^2  } \,,
\end{equation}
is given in terms of the Wilson coefficients by~\cite{Hiller:2014ula}
\begin{align}
R_K \simeq 1 +   \frac{2 C_{10}^{\mbox{\scriptsize{SM}}}    }{   |C_9^{\mbox{\scriptsize{SM}}}|^2 + |C_{10}^{\mbox{\scriptsize{SM}}}|^2  }   \left(      C_{10}^{\mbox{\scriptsize{NP}} \mu}  -      C_{10}^{\mbox{\scriptsize{NP}} e}      \right)    +    \frac{   |C_{10}^{\mbox{\scriptsize{NP}} \mu}|^2  - |C_{10}^{\mbox{\scriptsize{NP}} e}|^2   }{    |C_9^{\mbox{\scriptsize{SM}}}|^2 + |C_{10}^{\mbox{\scriptsize{SM}}}|^2    }  \,.
\end{align}
Here we have taken into account the absence of right-handed FCNCs in the down quark sector and the fact that $C_{9}^{\mbox{\scriptsize{NP}} \mu} = C_{9}^{\mbox{\scriptsize{NP}} e}$.    Deviations from the SM in $R_K$ can only be accommodated in our model from differences between $C_{10}^{\mbox{\scriptsize{NP}} e}$ and $C_{10}^{\mbox{\scriptsize{NP}} \mu}$.

\begin{figure}[!h]
\begin{center}
~
 \includegraphics[width=5cm]{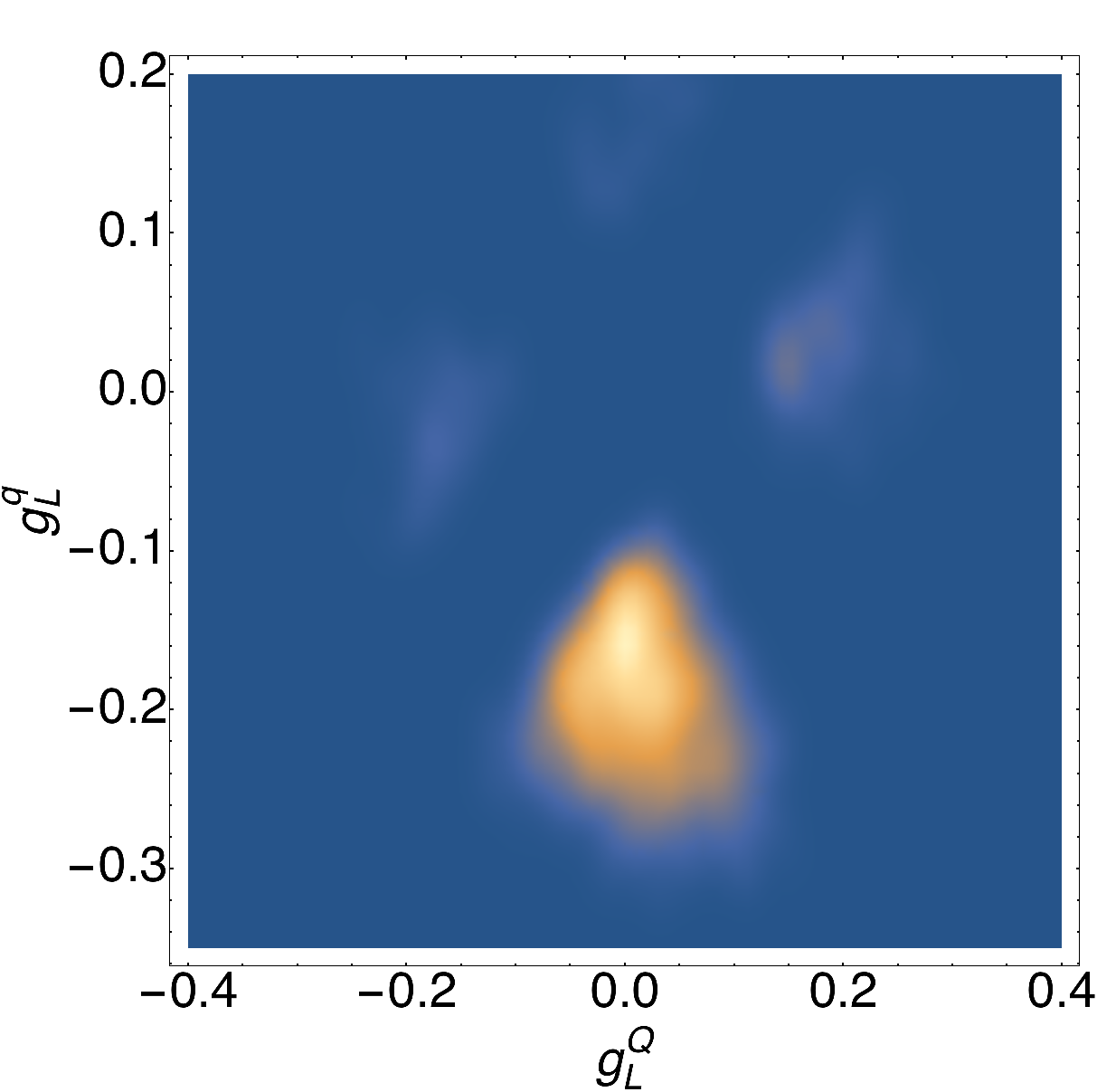}
 ~
 \includegraphics[width=5cm]{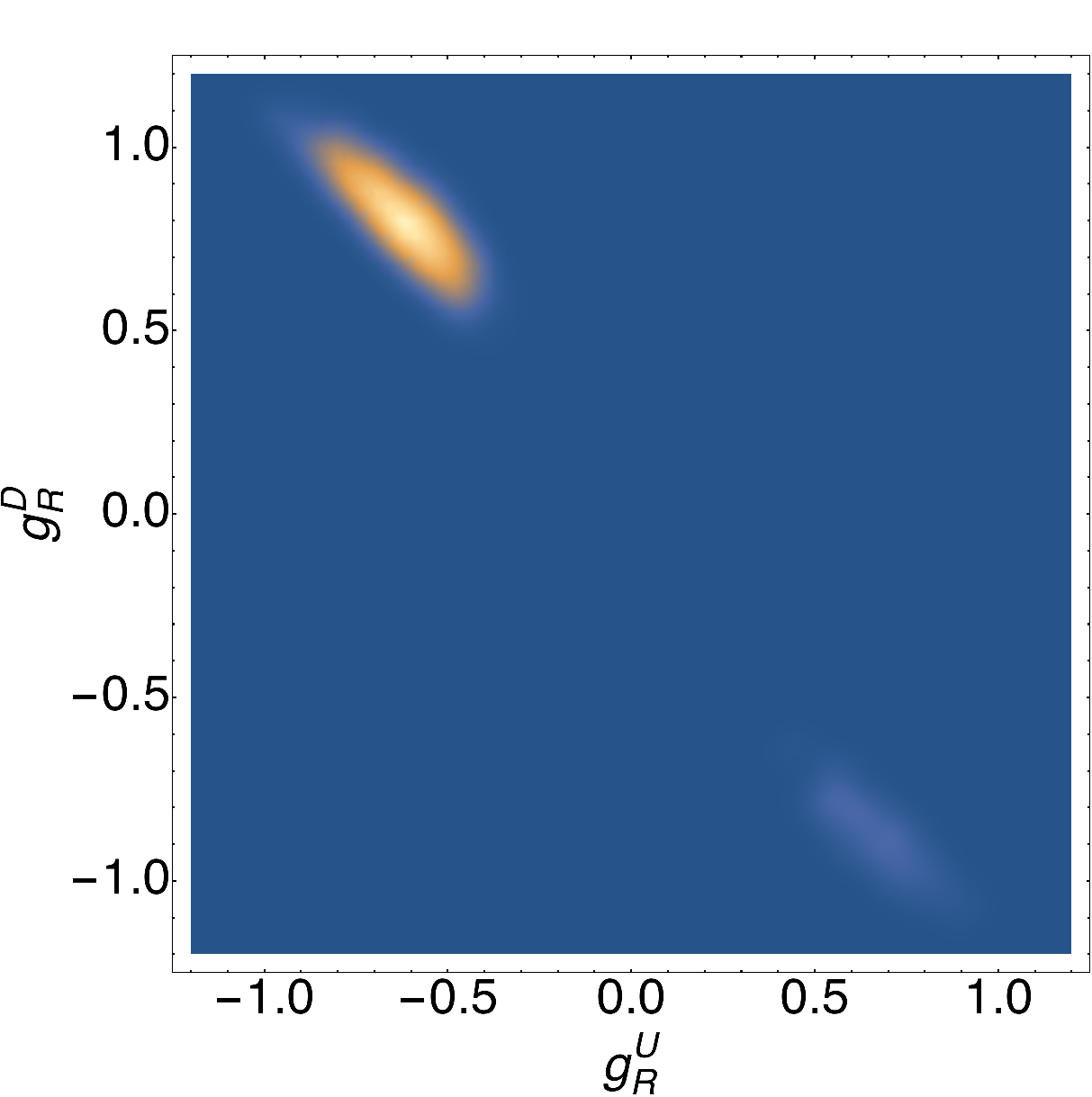}
~
  \includegraphics[width=5cm]{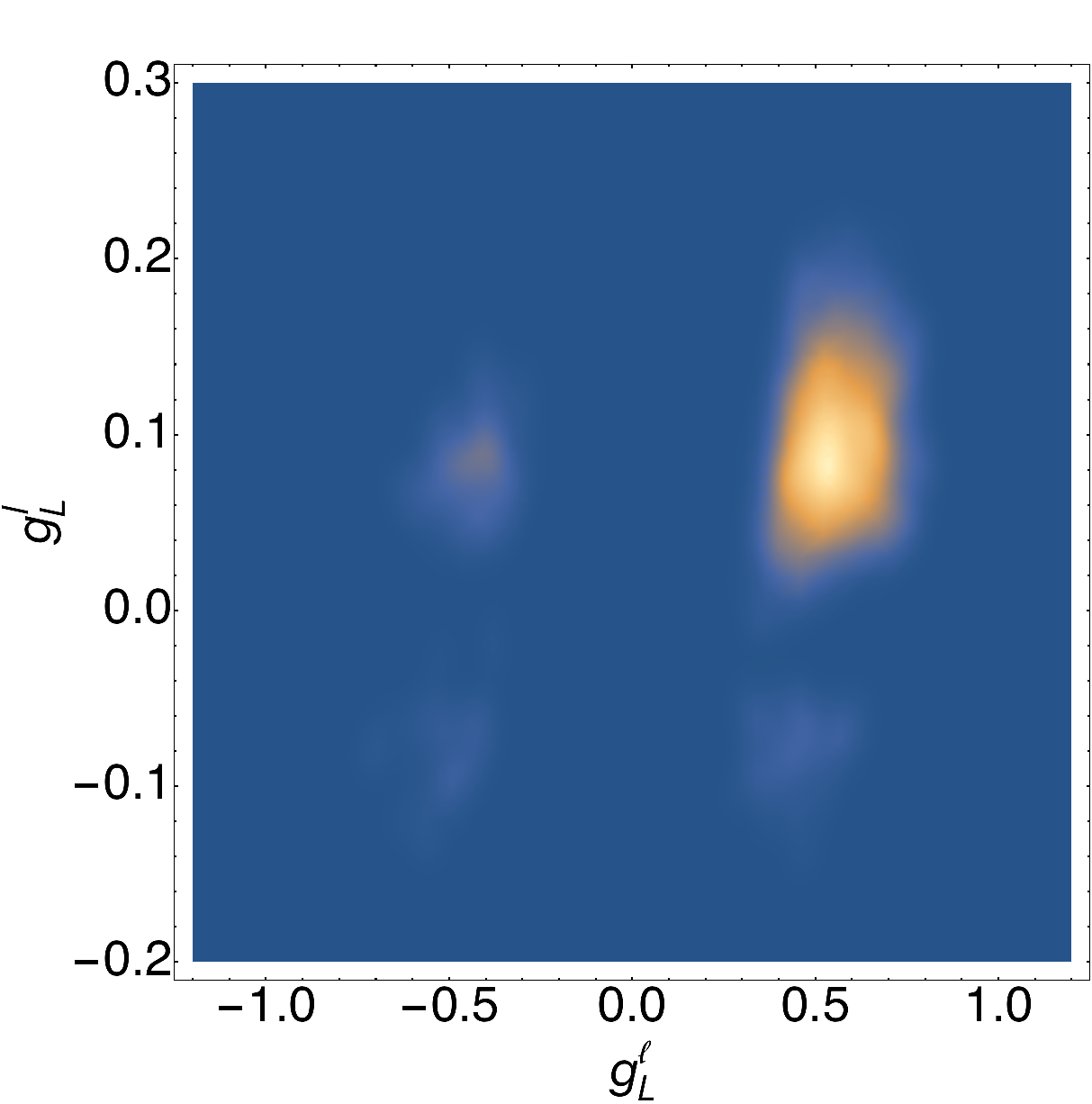}\\
  ~
  \includegraphics[width=5cm]{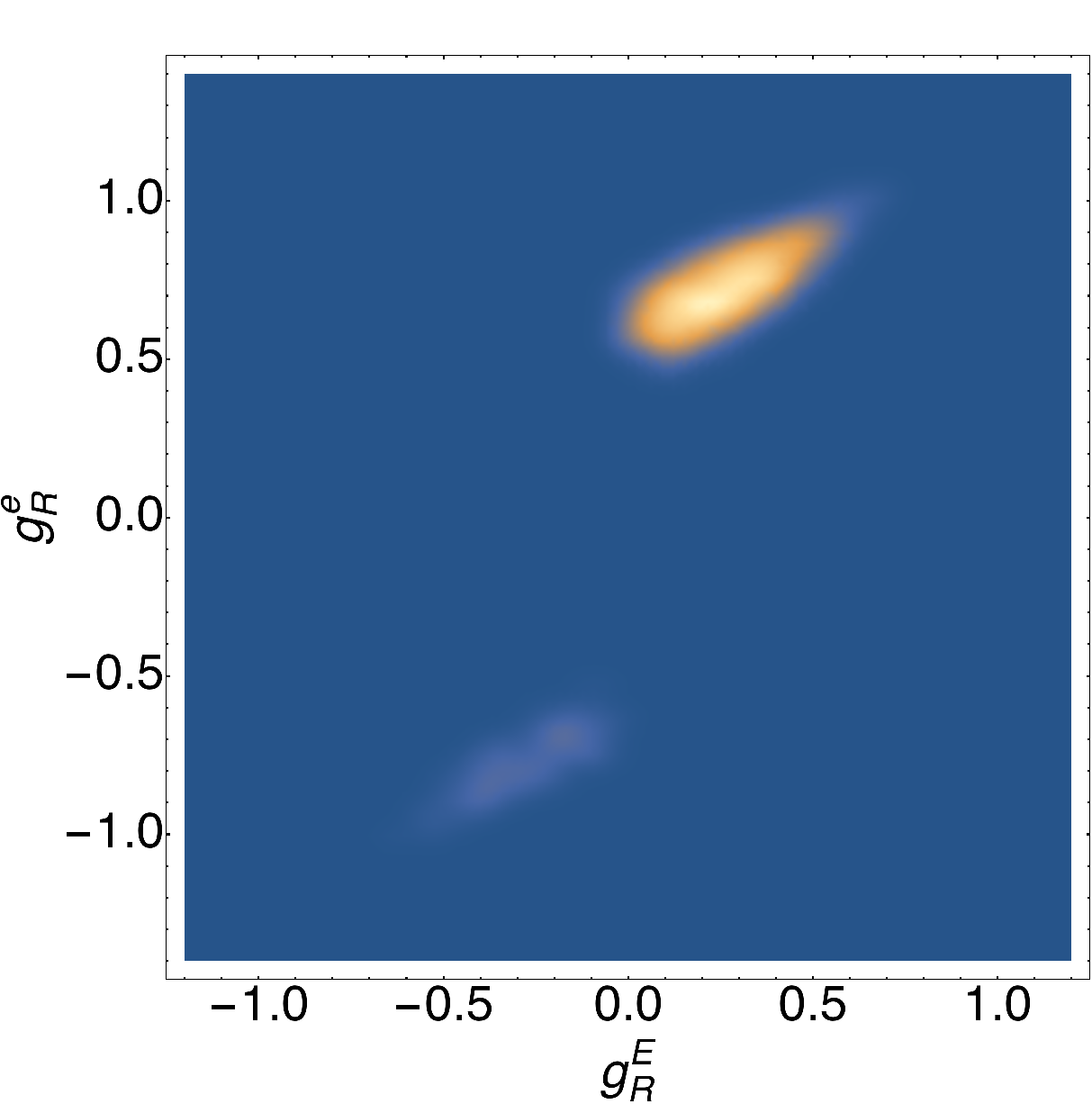}
  ~
   \includegraphics[width=5cm]{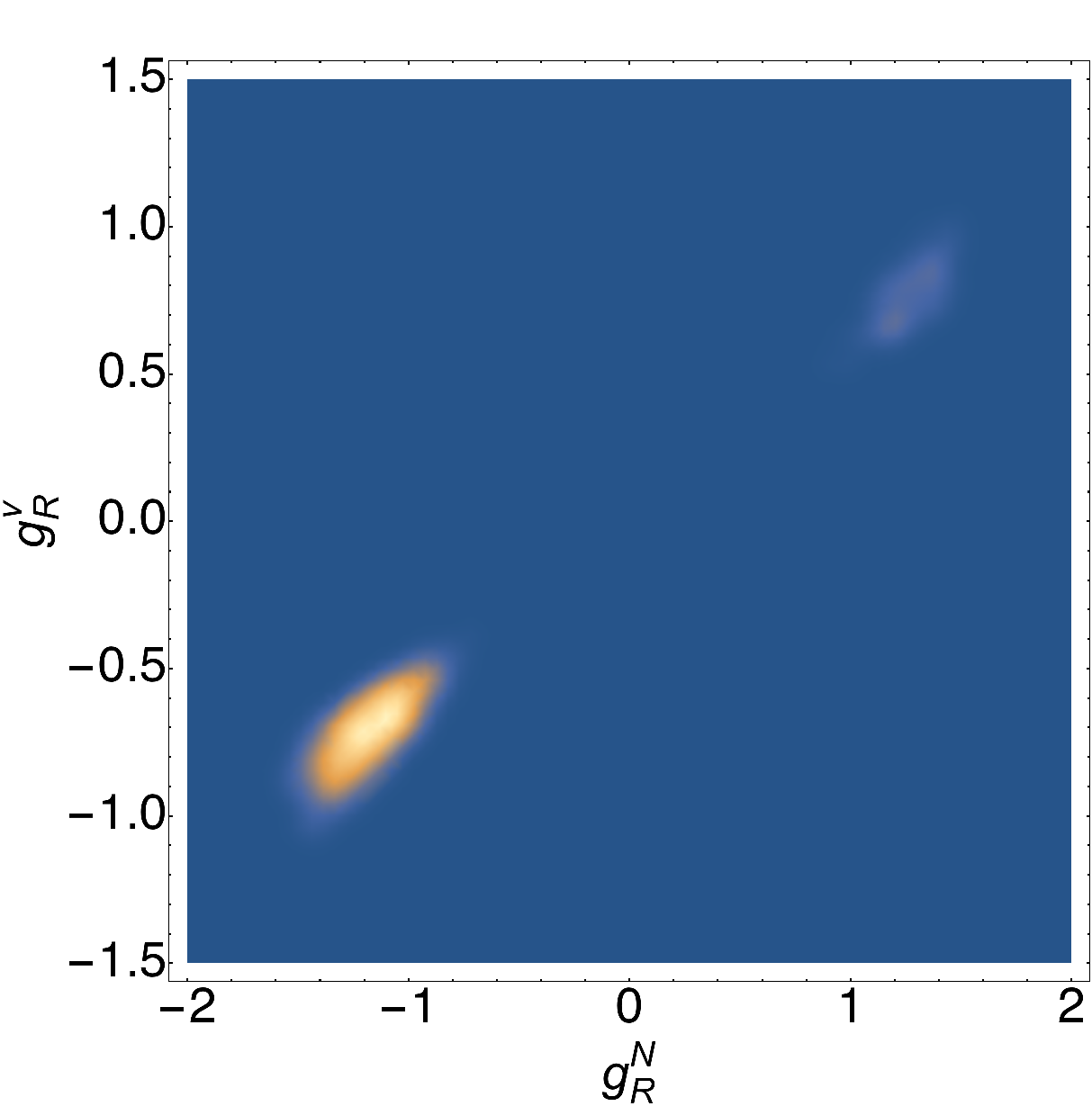}
 \end{center}
  \caption{\textit{Constraints on the couplings $g_{L,R}^{f}$ from $Z^{\prime}$ searches at the LHC, $B_s$-meson mixing, $b \rightarrow s \mu^+ \mu^-$ data and the $R_K$ measurement at $95\%$~CL.    The coupling $g_{L}^{q}$ ($g_{L}^{Q}$) determines the left-handed $Z^{\prime}$ couplings to the first two (third) quark generations.      Right handed $Z^{\prime}$ couplings to quarks are determined by $g_{R}^{U,D}$.     The $Z^{\prime}$ coupling to electrons is determined by $g_{L}^{l}$ and $g_{R}^{e}$ while those to the second and third lepton generations are determined by $g_{L}^{\ell}$ and $g_{R}^{E}$.    $g_{R}^{\nu}$ and $g_{R}^{N}$ represent the $Z^{\prime}$ couplings to right-handed neutrinos.   \label{figgco} }}
\end{figure}

We have performed a scan over the parameter space of our model searching for those combinations of parameters which can accommodate the observed anomalies in $b \rightarrow s \ell^+ \ell^-$ data.     We use the bounds from branching ratios and angular observables in $b \rightarrow s \mu^+ \mu^-$ decays provided in~\cite{Descotes-Genon:2015uva}.   We reconstruct the likelihood from the iso-contours of $\Delta \chi^2$ provided in the plane $(C_9^{\mbox{\scriptsize{NP}} \mu} ,  C_{10}^{\mbox{\scriptsize{NP}} \mu} )$, assuming these are well approximated by a bivariate normal distribution.   The measurement of $R_K$ is included in the analysis by modeling the LHCb measurement by a Gaussian distribution.\footnote{
Possible correlations between $R_K$ and the different observables included in the global fit of $b \rightarrow s \mu^+ \mu^-$ data performed in~\cite{Descotes-Genon:2015uva} are neglected here.}
We consider bounds from the measured mass difference in the $B_s$-meson system,  for which contributions as large as $20\%$ remain allowed at $95\%$~CL~\cite{Celis:2015ara}.        Constraints from $Z^{\prime}$ direct searches at the LHC in the di-muon and di-electron channels are also taken into account.  These are implemented through experimental bounds provided in the so-called $(c_d,c_u)$ plane~\cite{Accomando:2010fz,Khachatryan:2014fba}.  Allowed values within our model at $95\%$~CL from this analysis are shown in Figures~\ref{figWC} and \ref{figtu}, the more likely regions of the parameter space have a lighter shading.

We obtain a strong correlation between $C_{9}^{\mbox{\scriptsize{NP}} \mu}$ and $C_{10}^{\mbox{\scriptsize{NP}} e}$, with $C_{10}^{\mbox{\scriptsize{NP}} e}$ taking negative values in the range $\sim[-0.8,-0.3]$.     The value of $C_{10}^{\mbox{\scriptsize{NP}} \mu}$ is compatible with zero and shows preference for positive values due to $b \rightarrow s \mu^+ \mu^-$ data~\cite{Descotes-Genon:2015uva}. The mass of the $Z^{\prime}$ boson is bounded to lie in the range $\sim [3.5, 5.5]$~TeV, within the reach of the next LHC runs~\cite{Godfrey:2013eta}.    We find that the $Z^{\prime}$ would decay slightly more often to electrons than to muons, with $\mathrm{Br}(Z^{\prime} \rightarrow \mu^+\mu^-)/\mathrm{Br}(Z^{\prime} \rightarrow e^+e^-) \sim [0.5,0.9]$.     Strong bounds and correlations are obtained from our analysis for the couplings $g_{L,R}^{f}$ as shown in Figure~\ref{figgco}.

\section{$\mathrm{\mathbf{U(1)}}$ mass mixing and $\mathbf{Z^{\prime}}$ in intersecting brane models  \label{eqmasmix}}
In this section, we give a brief introduction to intersecting D-brane models as well as the mass mixing effect of the extra $\mathrm{U(1)}$'s.
In the literature, D-brane models were usually built with three or four stacks of
D-branes intersecting with each other and matter fields are realized at intersections of branes
transforming generically in the bi-fundamental representations of the gauge groups associated to the branes.
Additional $\mathrm{U(1)}$'s beyond the SM hypercharge appear naturally in intersecting D-branes constructions.
Each D-brane would give rise to a $\mathrm{U(1)}$ and $\mathrm{U(N)}$ arises from $N$ overlapping D-branes.
Thus instead of getting $\mathrm{SU(3)}\times \mathrm{SU(2)} \times \mathrm{U(1)}$ SM gauge group,
one would get $\mathrm{U(3)} \times \mathrm{U(2)} \times \mathrm{U(1)}$ and possibly with more $\mathrm{U(1)}$'s in the set-up.
The hypercharge appears as a linear combination of these $\mathrm{U(1)}$'s.

Extra $\mathrm{U(1)}$'s may also come from D-branes in the hidden sector,
which do not intersect with visible branes that give rise to the SM.
These hidden $\mathrm{U(1)}$'s are phenomenologically very interesting and
through mass mixing effect they could mix with $\mathrm{U(1)}$'s from the visible sector
and thus generate a portal between visible and hidden sectors,
which could explain the nature of dark matter, as was discussed in~\cite{Feng:2014eja,Feng:2014cla}.
These $\mathrm{U(1)}$'s in D-brane models are usually massive due to the coupling of
gauge fields to two-index antisymmetric tensors from closed string sector~\cite{Ibanez:1998qp,Poppitz:1998dj}.
These couplings are crucial for the generalized Green-Schwarz mechanism
which is responsible for the anomaly cancellation, and
at the same time they also generate a string scale St\"uckelberg mass to the $\mathrm{U(1)}$ gauge bosons.
Assuming a $\mathcal{O}(10)$~TeV string scale $M_S$ due to the large internal volume,
the stringy $Z'$s would be within the reach of the LHC,
and interesting phenomenology launches~\cite{Ghilencea:2002da,Langacker:2008yv,Berenstein:2008xg,Kumar:2007zza,Antoniadis:2009ze,Williams:2011qb,Anchordoqui:2011ag,Anchordoqui:2011eg,Feng:2014eja,Feng:2014cla,Anchordoqui:2015uea}.

We focus on the Type IIA string theory compactified on a Calabi-Yau threefold $\mathcal M$,
with the orientifold action $\Omega \bar\sigma$ where $\Omega$ denotes the world-sheet
parity transformation and $\bar \sigma$ is the anti-holomorphic involution on the compact six-dimensional space $\mathcal M$.
The homology group $H_3(\mathcal M)$ can be decomposed into its $\Omega \bar\sigma$
even and odd parts, $H_3(\mathcal M) = H_3^+(\mathcal M) \oplus H_3^-(\mathcal M)$.
We choose a basis in which $\alpha_i \in H_3^+(\mathcal M)$ and $\beta_i \in H_3^-(\mathcal M)$
with $i = 1,2,\cdots, l\equiv h^{2,1}+1$, such that
\begin{equation}
\alpha_{i}\cdot \beta^{j}=-\beta_{i}\cdot \alpha^{j}=\delta_{i}^{\phantom{i}j}\,,\qquad \alpha_i\cdot \alpha_j=\beta_i\cdot \beta_j=0\,.
\end{equation}
The orientifold O6-planes wrap 3-cycles $\Pi_{\rm O6}$,
and the Ramond-Ramond (RR) charges of the O6-planes are cancelled by $n$ stacks of D6-branes wrapping
on the three-cycles $\Pi_a$ and their orientifold images $\Pi_{a'}$ ($a=1,2,\cdots,n$).
These three-cycles can be expanded in the basis introduced above as
\begin{equation}
\Pi_{a}=S_{a}^{\phantom{a}i}\alpha_{i}+R_{a i}\beta^{i}\,,
\qquad \Pi_{a'}=S_{a}^{\phantom{a}i}\alpha_{i}-R_{a i}\beta^{i}\,,
\qquad \Pi_{\mathrm{O6}} = L^{i} \alpha_i\,,\label{EObasis}
\end{equation}
where the entries of matrix $S,R,L$ are all integers (or half integers when considering a tilted orientifold~\cite{Ibanez:2001nd}).

For a consistent brane model, tadpoles induced by the D6-brane are cancelled by orientifold 6-planes carrying $-4$ units of brane charge. Tadpole cancellation required three-cycles wrapped by branes and also $\mathrm{O6}$-plane to satisfy
\begin{equation}
\sum_a N_a \Pi_a + \sum_b N_b \Pi_{b'} - 4 \Pi_{\mathrm{O6}} = 0 \,,
\label{Tadpole}
\end{equation}
where $N_a$ is the number of overlapping branes in stack $a$ wrapping the three-cycle $\Pi_a$.

At the intersection of $a$ and $b$ brane stacks, chiral fermions are open strings stretching between these two stacks and transform under the bi-fundamental representation $(\square_a,\bar{\square}_b)$ with the corresponding $\mathrm{U(1)}$ charges $(+1,-1)_{a,b}$. The number of replicas of such chiral fermions, is given by the intersection number which presents
the number of times the wrapped cycles intersect with each other
\begin{equation}
I_{ab} = \Pi_a \cdot \Pi_b = S_{ai} R_{i}^{T\,b}-R_{ai}S_i^{T\,b}\,.
\end{equation}
In addition, the intersection number between a cycle and another cycle's orientifold image is given by
\begin{equation}
I_{ab'} = \Pi_a \cdot \Pi_{b'} = - S_{ai} R_{i}^{T\,b}-R_{ai}S_i^{T\,b}\,.
\end{equation}

For toroidal models, D6-branes wrap a three-cycle on a six-dimensional torus
which can be factorized into $T^2 \times T^2 \times T^2$,
and the D$6_a$-brane warps on the $i$-th $T^2$ with wrapping numbers $(n_a^i, m_a^i)$.
The intersection numbers of $a$ and $b$ brane stacks
as well as $a$ and $b'$ which is the orientifold image of stack $b$
are given by
\begin{align}
I_{ab} &= \Pi_a \cdot \Pi_b = (n_a^1 m_b^1 - m_a^1 n_b^1)(n_a^2 m_b^2 - m_a^2 n_b^2)(n_a^3 m_b^3 - m_a^3 n_b^3)\,,\\
I_{ab'} &= \Pi_a \cdot \Pi_{b'} = - (n_a^1 m_b^1 + m_a^1 n_b^1)(n_a^2 m_b^2 + m_a^2 n_b^2)(n_a^3 m_b^3 + m_a^3 n_b^3)\,.
\end{align}
The intersection numbers give rise to the number of families of the chiral fermions arising at the corresponding intersections.

On the other hand, the $B\wedge F$ couplings are obtained through the Kaluza-Klein reduction of D6-brane Chern-Simons action
\begin{equation}
S_{BF} = \frac{1}{2} \left( \int_{\Pi_a} C_5 \wedge \mathrm{tr} F_a
- \int_{\Pi_a'} C_5 \wedge \mathrm{tr} F_a \right) 
 = \sum_a N_a R_{a i} B_2^i \wedge F_a \,,
\end{equation}
where $B_2^i = \int_{\beta_i} C_5$ and $C_5$ is the RR 5-form.
The $B\wedge F$ couplings can then give rise to
\begin{equation}
\mathcal{L}_{\rm St} \sim \tfrac{1}{2} \mathcal{G}_{ij}
(\partial_\mu a_i + N_a R_{ai} A_{a,\mu})(\partial^\mu a_j+ N_b R_{bj} A^{\mu}_b)\,,
\end{equation}
where the matrix $\mathcal{G}_{ij}$ is the (positive-definite) metric of the complex structure moduli space,
and the RR axion $a_i = \int_{\alpha_i} C_3$ which is dual to $B_2^i$.
The gauge boson and axion couplings play a crucial role canceling the triangle anomalies of the anomalous $\mathrm{U(1)}$'s~\cite{Aldazabal:2000dg}.
$\mathrm{U(1)}$ gauge bosons then require St\"uckelberg mass and
the $\mathrm{U(1)}$ mass-squared matrix takes the form~\cite{Ghilencea:2002da,Feng:2014cla}
\begin{equation}
M_{ab}^{2} = g_a g_b N_a R_{ai} \mathcal{G}_{ij} N_b R^{T,ib} M_S^2\,,
\label{MASSM}
\end{equation}
where $M_S$ is the string scale,
the $\mathrm{U(1)}$ indices $a,b=1,2,\cdots,n$ runs over all the branes in the set-up,
$i,j=1,2,\cdots,h_{2,1}+1$ are the complex structure moduli indices,
and the entries of the $R$ matrix are all integers or half integers.
The complex structure moduli matrix $\mathcal{G}_{ij}$ for toroidal case was analyzed in~\cite{Feng:2014cla}.
In this work, to illustrate our idea we set this matrix to be the identity matrix for simplicity.

Now we discuss briefly the $\mathrm{U(1)}$ mass mixing effect in intersecting brane models.
Assuming no kinetic mixing, the Lagrangian of $n$ $\mathrm{U(1)}$ fields reads
\begin{align}
\mathcal{L} & =-\frac{1}{4}\sum_{a=1}^n F_{a}^{2}+\frac{1}{2}A_{a}M_{ab}^{2}A_{b}+\sum_{a=1}^n \bar{\psi}_{a}(i\slashed\partial+g_{a}q_{a}\slashed A_{a})\psi_{a}\,,
\end{align}
where $\psi_{a}$ denotes the matter fields charged under $\mathrm{U(1)}_{a}$.
An orthogonal matrix $O$ would bring $M^{2}$ into diagonal form
with its elements being the eigenvalues of $M^{2}$:
\begin{equation}
O^{T}M^{2}O={\rm{diag}}\{ \lambda_{1}^{2}, \lambda_{2}^{2},\cdots , \lambda_{n}^{2}
\}\equiv D^{2}\,,
\end{equation}
where the eigenvalues are sorted from small to large, i.e., $\lambda_{i}<\lambda_{j}$
for $i<j$. $\lambda_{1}=0$ corresponds to the mass of the hypercharge
gauge boson $B_{\mu}\equiv A_{1,\mu}^{(m)}$. We define the lightest
massive $\mathrm{U(1)}$ to be $\mathrm{U(1)}'$, and $Z'$ is the corresponding gauge boson.
In some models (including the model considered later),
there might be more zero's in $D^{2}$ in addition to the hypercharge,
and they should gain a mass at low energies.

The above transformation also takes the gauge fields from their original basis into the mass
(physical) eigenbasis, denoted by the upper index $^{(m)}$:
\begin{equation}
A_{i}^{(m)}=O_{ia}^{T}A_{a}\,.
\end{equation}
The column vectors of the orthogonal matrix $O$ are just the eigenvectors
of $M^{2}$. The first column vector $\vec{v}$ gives rise to the hypercharge combination
$\mathrm{U(1)}_Y = v_{1}\mathrm{U(1)}_{a}+v_{2}\mathrm{U(1)}_{b}+\cdots+v_{n}\mathrm{U(1)}_{n}$,
which is totally determined by the construction.
The smallest non-zero eigenvalue is the mass-square of the $Z'$ gauge boson,
and the corresponding column vector
$\vec{\zeta}=(\zeta_{1},\zeta_{2},\cdots,\zeta_{n})$
gives rise to the $Z'$
\begin{equation}
\mathrm{U(1)}'=\zeta_{1}\mathrm{U(1)}_{a}+\zeta_{2}\mathrm{U(1)}_{b}+\cdots+\zeta_{n}\mathrm{U(1)}_{n}\,,
\end{equation}
where the vector elements are determined by the details of the $\mathrm{U(1)}$ mass-squared matrix.

After the mass mixing, the Lagrangian in the $\mathrm{U(1)}$ gauge boson mass
eigenbasis reads
\begin{equation}
\mathcal{L}=-\frac{1}{4}\sum_{i=1}^n F_{i}^{(m)2}+\frac{1}{2}D_{ii}^{2}(A_{i}^{(m)})^{2}+\sum_{a=1}^n \bar{\psi}_{a}(i\slashed\partial+g_{a}q_{a}O_{ai}\slashed A_{i}^{(m)})\psi_{a}\,.
\end{equation}
Since the elements in the orthogonal matrix $O$ are in general irrational
numbers (except for the first column corresponding to the hypercharge, which are all fractional numbers by construction),
the gauge charges in the $\mathrm{U(1)}$ mass eigenbasis are not quantized.
For a matter field carrying $q_{a}$ under $\mathrm{U(1)}_{a}$ with the gauge
coupling $g_{a}$, after mass mixing it couples to the gauge field
$A_{i}^{(m)}$ in the mass eigenbasis with strength $g_{i}^{(m)}Q_{i}^{(m)}\equiv\sum_{a}g_{a}q_{a}O_{ai}$.
In intersecting brane set-ups, chiral fermions are usually realized
as bi-fundamental fields and hence charged under two gauge groups.

The gauge couplings on the visible branes can be identified with the
SM gauge coupling constants running to the string energy scale.
For the $\mathrm{U(3)}$ stack, $g_{a}=\frac{1}{\sqrt{6}}g_{{\rm QCD}}$;
and for the $\mathrm{U(2)}$ stack, $g_{b}=\frac{1}{2}g_{2}$. The hypercharge
gauge coupling yields
\begin{equation}
\frac{1}{g_{Y}^{2}}=\frac{v_{1}^{2}}{g_{a}^{2}}+\frac{v_{2}^{2}}{g_{b}^{2}}+\cdots+\frac{v_{n}^{2}}{g_{n}^{2}}\,.
\label{Hyper}
\end{equation}

Since the Higgs fields are also realized as open strings stretching
between two stacks of visible branes and hence are charged under the
two $\mathrm{U(1)}$'s. After the mass mixing, the Higgs fields would also
be charged under all other $\mathrm{U(1)}$'s in the mass eigenbasis, and couple
to all these massive $\mathrm{U(1)}$ gauge bosons. Hence after the electroweak
symmetry breaking, all the gauge boson mass would be corrected by the Higgs mechanism. The
covariant derivative reads
\begin{equation}
D_{\mu}=\partial_{\mu}-ig_{2}A_{\mu}^{a}T^{a}-i\frac{g_{Y}}{2}B_{\mu}-i\sum_{i} g_{i}^{(m)}Q_{i}^{(m)}A_{i}^{(m)}\,,
\end{equation}
where $T^{a}=\sigma^{a}/2$ is the $\mathrm{SU(2)}$ generator, and $B_{\mu}$ the hypercharge
gauge boson. The mass terms of all the $\mathrm{U(1)}$'s take the form
\begin{align}
\mathcal{L}_{m} & =D_{\mu}\phi D^{\mu}\phi+\frac{1}{2}D_{ii}^{2}(A_{i}^{(m)})^{2}\nonumber \\
 & =\frac{v^{2}}{8}\Big[g_{2}^{2}(A_{\mu}^{1})^{2}+g_{2}^{2}(A_{\mu}^{1})^{2}
+\big(g_{Y}B_{\mu}-g_{2}A_{\mu}^{3}+2\sum_{i} g_{i}^{(m)}Q_{i}^{(m)}A_{i}^{(m)}\big)^{2}\Big]+\frac{1}{2}D_{ii}^{2}(A_{i}^{(m)})^{2}\,.\label{LM}
\end{align}
$A_{\mu}^{1}$ and $A_{\mu}^{2}$ give rise to $W^{\pm}$ and the
mass mixing only occurs within $A_{\mu}^{3},A_{i}^{(m)}$. One needs
to perform another diagonalization to determine the final mass eigenstates
of all the $\mathrm{U(1)}$ gauge bosons. The special form of Eq.~\eqref{LM} ensures
there is only one massless eigenstate $A_{\mu}^{\gamma}=(g_{Y}A_{\mu}^{3}+g_{2}B_{\mu})/(g_{2}^{2}+g_{Y}^{2})^{1/2}$
which will be identified to be the photon. As one can see, the photon
does not couple to any hidden matter, and thus all the hidden fields are
exactly electrically neutral. The electric charge remains unchanged,
i.e., $e=g_{2}g_{Y}/(g_{2}^{2}+g_{Y}^{2})^{1/2}$. The
$Z$ boson would be a mixture of $A_{\mu}^{3}$ and all the $A_{i}^{(m)}$,
and its mass would receive a small correction due to mass mixing effects $M_{Z}=   (g_{2}^{2}+g_{Y}^{2})^{1/2} v/2  +\mathcal{O}\left(v^{2}/M_{S}^{2}\right)$~\cite{Ghilencea:2002da,Coriano':2005js}.
Since $v \ll M_S$,
the mass of the extra $\mathrm{U(1)}$'s comes mainly from the St\"uckelberg mechanism and will be around the string
scale.

A phenomenologically interesting intersecting D-brane
model usually requires a low string scale, in which case the spacetime
extends into large extra dimensions~\cite{Antoniadis:1997zg,Antoniadis:1998ig}.
Brane models with a low string scale posses low lying string excitations,
which would be clear signal for string theory at the LHC collider~\cite{Cullen:2000ef,Burikham:2004su,Lust:2008qc,Anchordoqui:2008di,Carmi:2011dt,Anchordoqui:2014wha}.
A low string scale can be achieved by enlarging some of the transverse compactification
radii to the D-branes where the SM is located. Since the D6-brane gauge coupling
constants are proportional to the inverse of the volumes of the wrapped
3-cycles, at least some of the cycles (especially the ones that visible
branes wrapped on) has to be kept small compared to the overall volume.
For the toroidal model in type IIA string theory, one could think
the six-torus still be kept small but connect to a large volume manifold.
There is another possibility discussed in~\cite{Feng:2014eja,Feng:2014cla}
that one can include a hidden sector with many hidden branes which do not intersect
with the visible branes that realize the SM, while
the hidden $\mathrm{U(1)}$'s could still mix with the visible $\mathrm{U(1)}$'s via the $\mathrm{U(1)}$ mass-squared matrix.
In this set-up, one has the possibility to generate a light $Z'$ with its mass several
orders lower than the fundamental string scale,
due to the significant repulsion effect to the eigenvalues of a large (semi)positive-definite matrix with entries of the same order.

Despite gaining a St\"ukelberg mass, these anomalous $\mathrm{U(1)}$'s (for example $\mathrm{U(1)}_B$) remain unbroken
at the perturbative level in the low energy effective theory~\cite{Ibanez:1999it},
and can thus protect the stability of the proton.
However, D-brane instanton effects may break these symmetries and allow for baryon number violating couplings.
This may be cured by the implementation of discrete gauge symmetries~\cite{BerasaluceGonzalez:2011wy,Anastasopoulos:2012zu,Honecker:2013sww} which again forbid these unwanted couplings.

\section{Intersecting brane model for $b \rightarrow s \ell^+ \ell^-$ anomalies}
\label{sec:anomaM}

In intersecting brane models, chiral fermions are usually realized as open strings
stretching between two stacks of D-branes, and thus carry gauge charges
under two $\mathrm{U(1)}$'s in the original D-brane basis. After the $\mathrm{U(1)}$
mass mixing, the chiral fermions would couple to the $Z'$ with the
strength $\sum_{a}g_{a}q_{a}\zeta_{a}$ where $a$ runs over the D-branes
they are attaching to, and $\vec{\zeta}$ is the eigenvector of the
$\mathrm{U(1)}$ mass-squared matrix corresponding to the $Z'$ gauge boson.
Given the D-brane construction, the gauge couplings of the
$Z'$ to fermions $g_{L,R}^{f}$ in Eq.~\eqref{eqlagz} are determined by $g^f_{L,R} \equiv\sum_{a}g_{a}q_{a L,R}^{f} \,\zeta_{a}$.

The $\mathrm{U(1)}'$ gauge charges of the SM fermions are very model dependent.
For $\mathrm{USp(2)}$ type D-brane model constructions proposed in~\cite{Ibanez:2006da},
the $Z'$ couplings to the quark sector are family-universal.
While for $\mathrm{U(2)}$-type intersecting D-brane models~\cite{Ibanez:2001nd}, the $\mathrm{U(1)}'$ charges of left-handed quarks are family non-universal.
We focus on the $\mathrm{U(2)}$-type intersecting D-brane models
since they can generate FCNCs in the left-handed quark sector.
In addition, we also need that the $Z'$ couples differently to electrons and muons.

We consider a five-stack toroidal model in which two generations of leptons arise from the intersection between stack $c$ and $d$ branes,
while the other generation arise from the intersection between $c$ and an additional $e$ stack~\cite{Kokorelis:2002zz}.
Since the three generations of leptons are realized at different brane intersections,
this construction allows automatically family-dependent $Z'$ couplings to leptons.
For four-stack models, requiring that
different families of quarks and leptons are simultaneously realized at different brane intersections,
it seems necessary to add chiral exotics in the particle spectrum (e.g., fermions attached to the $\mathrm{U(2)}$ stack).
We will not discuss the four-stack set-ups in this paper.

The toroidal model we consider here is non-supersymmetric
and it was shown in~\cite{Ibanez:2001nd} that there are regions in the complex structure parameter space that the configuration is stable.
Since the main focus of this work is to
illustrate how the D-brane models can explain the $b \to s \ell^+ \ell^-$ anomalies,
we will skip this discussion here.

The five-stack intersecting D-brane model is anomaly-free and has the following wrapping numbers on $T^2 \times T^2 \times T^2$,
\begin{align*}
N_{a}=3\qquad & (1/\beta^{1},0)(n_{a},\epsilon\beta^{2})(3,\tilde{\epsilon}/2)\,,\\
N_{b}=2\qquad & (n_{b},-\epsilon\beta^{1})(1/\beta^{2},0)(\tilde{\epsilon},1/2)\,,\\
N_{c}=1\qquad & (n_{c},\epsilon\beta^{1})(1/\beta^{2},0)(0,1)\,,\\
N_{d}=1\qquad & (1/\beta^{1},0)(n_{d},2\epsilon\beta^{2})(1,-\tilde{\epsilon}/2)\,,\\
N_{e}=1\qquad & (1/\beta^{1},0)(n_{e},\epsilon\beta^{2})(1,-\tilde{\epsilon}/2)\,,
\end{align*}
where $\epsilon=\pm 1$, $\tilde \epsilon=\pm 1$, $\beta^i = 1-b^i$ is the Neveu-Schwarz background parameter and $b^i = 0, 1/2$,
and $n_a, n_b, n_c, n_d, n_e$ are five additional integer parameters.
One could get the following particle spectrum
\begin{align*}
\begin{aligned}
Q_L\qquad & ab\;\quad 1(3,2)\quad(+1,-1,0,0,0;+\tfrac{1}{6})\\
q_L\qquad & ab'\quad 2(3,2)\quad(+1,+1,0,0,0;+\tfrac{1}{6})\\
U_R\qquad & ac\;\quad 3(\bar{3},1)\quad(-1,0,+1,0,0;-\tfrac{2}{3})\\
D_R\qquad & ac'\quad 3(\bar{3},1)\quad(-1,0,-1,0,0;+\tfrac{1}{3})\\
\ell_L\qquad & bd\;\quad 2(2,1)\quad(0,-1,0,+1,0;-\tfrac{1}{2})\\
l_L\qquad & be\;\quad 1(2,1)\quad(0,-1,0,0,+1;-\tfrac{1}{2})\\
E_R\qquad & cd'\quad 2(1,1)\quad(0,0,-1,-1,0;+1)\\
N_R\qquad & cd\;\quad 2(1,1)\quad(0,0,+1,-1,0;\ 0\ )\\
e_R\qquad & ce'\quad 1(1,1)\quad(0,0,-1,0,-1;+1)\\
\nu_R\qquad & ce\;\quad 1(1,1)\quad(0,0,+1,0,-1;\ 0\ )
\end{aligned}
\end{align*}
where $Q_L$ arises from the intersection of  $a,b$ stacks of D-branes,
$q_L$ is realized from the intersection of  $a$ stack and the orientifold image of $b$ stack of branes,
and similar for other fields.
The charges in the parentheses show the $\mathrm{U(1)}$ charges for $a,b,c,d,e$ stack of branes and the hypercharge $\mathrm{U(1)}_Y$ respectively.
Here $q,Q$ represent left-handed quarks realized at different intersections,
$U,D$ are right-handed up and down type quarks,
$(l,\ell)_L$, $(e,\nu,E,N)_R$ are left-handed and right-handed leptons arising from different brane intersections.
The hypercharge is given by
\begin{equation}
\mathrm{U(1)}_Y = \frac{1}{6} \mathrm{U(1)}_a -\frac{1}{2} \mathrm{U(1)}_c -\frac{1}{2} \mathrm{U(1)}_d -\frac{1}{2} \mathrm{U(1)}_e\,.
\end{equation}
This combination imposes the following condition
\begin{equation}  \label{eqconn1}
n_c = \frac{\tilde \epsilon \beta^2}{2 \beta^1} (n_a + n_d +n_e)\,.
\end{equation}
In addition, the tadpole condition in Eq.~\eqref{Tadpole} gives another constraint
\begin{equation}   \label{eqconn2}
\frac{9n_a}{\beta^1} + \frac{2 n_b}{\beta^2} + \frac{n_d+n_e}{\beta^1} = 16 \,.
\end{equation}
One could relax this constraint by adding hidden branes
which do not intersect with the given branes and do not contribute to the rest of the tadpoles.
When adding $N_h$ hidden branes $(n_h^1,0)(n_h^2,0)(n_h^3,m_h)$ on $T^6$,
the new tadpole condition becomes
\begin{equation}
\frac{9n_a}{\beta^1} + \frac{2 n_b}{\beta^2} + \frac{n_d+n_e}{\beta^1} + N_h n_h^1 n_h^2 n_h^3= 16\,,
\end{equation}
which is a weak constrain.

In this set-up there is an extra anomaly-free $\mathrm{U(1)}$ which is orthogonal to the hypercharge
\begin{equation}
\mathrm{U(1)}'' =
\frac{3 \tilde \epsilon  \beta^2}{2 \beta^1} \big[
\mathrm{U(1)}_a - 3\mathrm{U(1)}_d -3 \mathrm{U(1)}_e
\big] + 19n_c \mathrm{U(1)}_c\,,
\end{equation}
which should acquire a mass at low energies.
However, this anomaly-free $\mathrm{U(1)}$ does not generate flavor-changing currents in either quark or lepton sector and thus we will not focus on this $\mathrm{U(1)}$ in this work.

Now we focus on the lightest massive $\mathrm{U(1)}$ in this model, i.e., $\mathrm{U(1)}'$ with the corresponding gauge boson $Z'$.   The couplings of $Z'$ to the SM chiral fermions read
\begin{gather}
 g_{L}^{q}=g_{a}\zeta_{a}+g_{b}\zeta_{b} \,,\qquad  g_{L}^{Q}=g_{a}\zeta_{a}-g_{b}\zeta_{b}\,,\\
g_{R}^{U}=-g_{a}\zeta_{a}+g_{c}\zeta_{c}\,,\qquad g_{R}^{D}=-g_{a}\zeta_{a}-g_{c}\zeta_{c}\,,\\
g_{L}^{l}=-g_{b}\zeta_{b}+g_{e}\zeta_{e}\,,\qquad   g_{L}^{\ell}=-g_{b}\zeta_{b}+g_{d}\zeta_{d} \,, \\
g_{R}^{e}=-g_{c}\zeta_{c}-g_{e}\zeta_{e}\,,\qquad  g_{R}^{E}=-g_{c}\zeta_{c}-g_{d}\zeta_{d}  \,, \\
 g_{R}^{\nu}=g_{c}\zeta_{c}-g_{e}\zeta_{e} \,,\qquad  g_{R}^{N}=g_{c}\zeta_{c}-g_{d}\zeta_{d} \,.
\end{gather}
We recall that
$q,Q$ represent left-handed quarks realized at different intersections,
$U,D$ are right-handed up and down type quarks,
$(l,\ell)_L$, $(e,\nu, E, N)_R$ are left-handed and right-handed leptons arising at different brane intersections.   We identify $q$ to be $(u,d)_L,(c,s)_L$, and $Q$ to be $(t,b)_L$;
we also identify $l$ to be $(e,\nu_e)_L$ and $\ell$ to be $(\mu,\nu_\mu)_L,(\tau,\nu_\tau)_L$.
$U,D$ represent $(u,c,t)_R$ and $(d,s,b)_R$ respectively.
$(e,\nu)_R$ is identified with the right-handed electron and electron-neutrino
while $(E, N)_R$ are identified with the second and third generation right-handed leptons.

The vector $\vec{\zeta}$ is obtained from diagonalizing the $\mathrm{U(1)}$ mass-squared matrix in Eq.~\eqref{MASSM},
thus it is completely determined by the internal geometry and intersecting brane construction.
The components of $\vec{\zeta}$ are in general irrational numbers
and hence particles charged under the massive $\mathrm{U(1)}$'s do not have quantized charges.
Recall the gauge couplings $g_{a}=\frac{1}{\sqrt{6}}g_{QCD}$,
$g_{b}=\frac{1}{2}g_{2}$ and Eq.~\eqref{Hyper} gives the hypercharge coupling
\begin{equation}  \label{eqYc}
g_Y^{-2} = \frac{1}{36}g_a^{-2}+ \frac{1}{4}g_c^{-2}+ \frac{1}{4}g_d^{-2}+\frac{1}{4}g_e^{-2}\,,
\end{equation}
which restricts the gauge couplings $g_{c,d,e}$.

Now we discuss briefly the Higgs sector of the model.
The Higgs fields are realized at the $b,c$ intersections and there could be two kinds of Higgs doublet~\cite{Ibanez:2001nd,Kokorelis:2002zz}
\begin{align*}
\begin{aligned}
h_1\qquad & bc\;\ \quad n_h(2,1) \,\quad (0,0,+1,-1,0;+\tfrac{1}{2})\\
h_2\qquad & bc\;\ \quad n_h(2,1) \,\quad (0,0,-1,+1,0;-\tfrac{1}{2})\\
H_1\qquad & bc'\;\quad n_H(2,1) \quad (0,0,-1,-1,0;+\tfrac{1}{2})\\
H_2\qquad &bc'\;\quad n_H(2,1) \quad (0,0,+1,+1,0;-\tfrac{1}{2})
\end{aligned}
\end{align*}
Here we recall our notation that
the charges in the parentheses show the $\mathrm{U(1)}$ charges for $a,b,c,d,e$ stack of branes and the hypercharge $\mathrm{U(1)}_Y$ respectively.
The number of the Higgs doublets are given by
\begin{equation}
n_{h^\pm} = I_{bc} = |\epsilon \beta_1 (n_b + n_c)|\,,\qquad
n_{H^\pm} = I_{bc'} = |\epsilon \beta_1 (n_b - n_c)|\,.
\label{HiggsN}
\end{equation}
Because of the  $\mathrm{U(1)}$ charges, the allowed Yukawa couplings in this model read~\cite{Kokorelis:2002zz}
\begin{align}
\mathcal{L}_{\rm Yukawa} &=
Y^t Q U_3 h_1 + Y^b Q D_3 H_2 + Y^U_{ij} q_i U_j H_1 + Y^D_{ij} q_i D_j h_2\nonumber\\
& +Y^\nu l \,\nu h_1 +Y^e l\, e H_2 + Y^N_{k_1 k_2} \ell_{k_1} N_{k_2} h_1 + Y^E_{k_1 k_2} \ell_{k_1} E_{k_2} H_2 + h.c.\,,
\end{align}
where $i = 1,2, j = 1,2,3, k_1,k_2 = 2,3$, and $Y$'s are the Yukawa coupling constants.
These Yukawa couplings constrain the unitary matrices $B^{d_L}$ and $B^\ell$ to be in the form as written in Eqs.~\eqref{BdL} and~\eqref{Bl}.
In our analysis, we require the number of both $H^\pm$ and $h^\pm$ Higgs doublets to be greater or equal to one.   For the phenomenological analysis presented in Section~\ref{sec:san} we have varied the wrapping numbers $n_{c,b}$ within $[-10,10]$ and $n_{d,e}$ within $[-30,30]$, $n_{a}$ is obtained as integer solutions to Eq.~\eqref{eqconn1}.          The gauge couplings $g_{c,d}$ are varied within $[0.3,1.2]$ with $g_{e}$ being obtained from the hypercharge condition~\eqref{eqYc}.     The string scale $M_S$ has been taken in the range $[10,20]$~TeV.  The matrix element $(   W_{d_L}^{\dag}   X_{d_L} )_{sb}$ in Eq.~\eqref{eqBq} was varied within $[10^{-3}, 0.2]$.       We have also studied the impact of imposing additionally the tadpole condition \eqref{eqconn2}.  We found that the latter condition has a small impact regarding the results presented in Section~\ref{sec:san}, though the values required for the wrapping numbers $n_{b,c}$ are larger.

\section{Conclusions \label{conc}}
Recent deviations from the SM observed in $b \rightarrow s \ell^+ \ell^-$ decays have triggered recent attention from the theoretical community as possible hints of new physics.  A possible new physics scenario is a heavy $Z^{\prime}$ boson with non-universal couplings to leptons and flavor changing couplings in the down-quark sector.     We have explored the possibility of realizing this scenario with a string inspired abelian gauge boson.    Such neutral gauge boson at the TeV scale arises
from intersecting D-brane models with a low string scale, in which case the spacetime extends into large extra dimensions.
We consider the five-stack model studied in~\cite{Kokorelis:2002zz}, for which left-handed quarks and leptons simultaneously arise from different D-brane intersections.     We find that this class of models can accommodate current anomalies in $b \rightarrow s \ell^+ \ell^-$ data in certain regions of the parameter space.

Future experimental prospects regarding $b \rightarrow s \ell^+ \ell^-$ transitions as well as direct $Z^{\prime}$ searches at the LHC make this scenario very appealing.     We find that the stringy $Z^{\prime}$ boson considered has non-negligible couplings to the first two quark generations and has a mass in the range $\sim [3.5, 5.5]$~TeV, so it should be possible to discover such state directly during the next LHC runs via Drell-Yan production in the di-electron or di-muon decay channels.    We find $\mathrm{Br}(Z^{\prime} \rightarrow \mu^+\mu^-)/\mathrm{Br}(Z^{\prime} \rightarrow e^+e^-) \sim [0.5,0.9]$ so that the $Z^{\prime}$ boson would decay slightly more to electrons than to muons.        Correlations arising in this model can be also tested with precise measurements of $b \rightarrow s \ell^+ \ell^-$ observables.     The model considered predicts $C_{9}^{\mbox{\scriptsize{NP}} \mu} = C_{9}^{\mbox{\scriptsize{NP}} e} $. This has important implications given that $C_{9}^{\mbox{\scriptsize{NP}} \mu} \sim -1$ is required to accommodate $b \rightarrow s \mu^+ \mu^-$ data.  Deviations from the SM in $R_K$ are explained due to differences between $C_{10}^{\mbox{\scriptsize{NP}} \mu}$ and $C_{10}^{\mbox{\scriptsize{NP}} e}$, given that there are no right-handed flavor changing currents in the down-quark sector.    We find $C_{10}^{\mbox{\scriptsize{NP}} e}\sim[-0.8,-0.3]$ while $C_{10}^{\mbox{\scriptsize{NP}} \mu}$ is compatible with zero, showing some preference for positive values.

\section*{Acknowledgments}
The work of A.C. is supported by the Alexander von Humboldt Foundation.
WZ.F. is supported by the Alexander von Humboldt Foundation and Max--Planck--Institut f\"ur Physik, M\"unchen.
A.C. and WZ.F. would like to thank the Study Tour 2015 organized by the Humboldt Foundation where the collaboration was started.
WZ.F. is also grateful to I\~{n}aki Garc\'{i}a-Etxebarria and Diego Regalado for helpful discussions.
The work of WZ.F and D.L. was supported by  the ERC Advanced Grant 32004 ``Strings and Gravity" and also by TRR 33.


\end{document}